\documentclass[useAMS,usenatbib]{mn2e}
\usepackage{epsfig}
\usepackage{amssymb}

\newcommand{\be}{\begin{equation}}
\newcommand{\ee}{\end{equation}}
\newcommand{\msun}{\mbox{M$_{\odot}$}}
\newcommand{\lsun}{\mbox{$L_{\odot}$}}
\newcommand{\msunyr}{\mbox{${\rm M}_{\odot}$ {\rm yr}$^{-1}$}}

\title[The effect of galaxy mass ratio on merger--driven starbursts]
{The effect of galaxy mass ratio on merger--driven starbursts}

\author[Cox et al.]
{T. J. Cox$^1$\footnotemark[1], Patrik Jonsson$^2$, Rachel S. Somerville$^3$,
        Joel R. Primack$^2$, and Avishai Dekel$^{4}$ \\
        $^1$Harvard-Smithsonian Center for Astrophysics, 60 Garden St., 
                Cambridge, MA 02138, USA\\
        $^2$Department of Physics, University of California, Santa Cruz, 
                1156 High St., Santa Cruz, CA 95064, USA\\
        $^3$Max-Planck-Institut f\"ur Astronomie,
                K\"onigstuhl 17, D-69117 Heidelberg, Germany\\
        $^4$Racah Institute of Physics, The Hebrew University, Jerusalem, 91904, Israel\\
        }

\begin{document}

\maketitle

\begin{abstract}
We employ numerical simulations of galaxy mergers to explore the effect of galaxy
mass ratio on merger--driven starbursts.  Our numerical simulations include radiative
cooling of gas, star formation, and stellar feedback to follow the interaction and
merger of four disk galaxies.  The galaxy models span a factor of 23 in total mass and
are designed to be representative of typical galaxies in the local Universe.  We find
that the merger--driven star formation is a strong function of merger mass ratio,
with very little, if any, induced star formation for large mass ratio mergers.  We
define a burst efficiency that is useful to characterize the merger--driven star
formation and test that it is insensitive to uncertainties in the feedback
parameterization.  In accord with previous work we find that the burst efficiency 
depends on the structure of the primary galaxy.  In particular, the presence of a 
massive stellar bulge stabilizes the disk and suppresses merger--driven star
formation for large mass ratio mergers.  Direct, co--planar merging orbits produce
the largest tidal disturbance and yield that most intense burst of star formation.
Contrary to naive expectations, a more compact distribution of gas or an increased
gas fraction both decrease the burst efficiency.  Owing to the efficient feedback
model and the newer version of SPH employed here, the burst efficiencies of the
mergers presented here are smaller than in previous studies.
\end{abstract}

\renewcommand{\thefootnote}{\fnsymbol{footnote}}
\footnotetext[1]{E-mail: tcox@cfa.harvard.edu}%

\section{Introduction}
\label{sec:intro}

While mergers between galaxies of approximately equal mass, so called major mergers, 
garner a significant amount of observational and theoretical interest, it is likely
that galaxies accreting much smaller objects are also important drivers of galaxy evolution.
This possibility is a natural consequence of the hierarchical growth of structure
predicted by the Lambda cold dark matter ($\Lambda$CDM; a.k.a., ``Double Dark'')
cosmology \citep[see, e.g.,][]{SPF}.  In this scenario
every dark matter halo, and presumably the galaxy it hosts, is assembled though a
continuous and varied merger history comprised of {\it many} small accretion events
and a {\it few} ``major'' ones \citep[see, e.g.,][]{LC93,We02}.

The transformation of galactic systems by major mergers has been well
studied.  Numerical simulations demonstrate that major mergers between
spiral galaxies produce remnants that generally resemble elliptical
galaxies, whose shapes are determined by random motions, i.e., they
are pressure-supported, and which possess projected mass distributions
that scale approximately as $\sim r^{1/4}$
\citep[e.g.,][]{BH91,BH92Rev,HRemI, HRemII,NB03,Cox06rot,Cox06}.
Observations of fine structure, kinematic subsystems, the bimodal
distributions of global clusters in elliptical galaxies, and the light
profiles of merger remnants lend support to the assertion that
elliptical galaxies are the byproduct of major galaxy mergers - in
what is commonly termed the ``merger hypothesis'' \citep{T77}.  For
further information regarding the observational and theoretical
evidence supporting the merger hypothesis, or the general dynamics of
galaxy interactions, we refer the interested reader to reviews by
\citet{BH92Rev}, \citet{Schweizer98}, and \citet{Stck06}.

While major mergers are likely to be the most dramatic events in a galaxy's
assembly history, it is likely that the more prevalent minor mergers give rise to
the wide variety of galaxy morphology that defines many classification systems
\citep{Hub26,deV77,vdB60,San75}.  The morphological transformation owing to the effects
of minor merger has been studied via numerous numerical simulations, which 
demonstrate that galactic disks become warped and heated as a result of the 
accretion of a satellite \citep{QG86,TO92, Qui93,WMH96,HC97,VW99,Fo01}, that the
satellite itself is tidally stripped to produce stellar streams throughout the galactic
halo \citep{Jo99,HW01,May02,BK07}, and that the remnant galaxy will be a systematically
earlier Hubble type \citep{ABP01,NB03,BCJ04,BJC05,EM06,Mal06}.  Furthermore, there is an
increasing amount of evidence that these theoretical expectations are consistent with
observations in the local universe \citep[e.g.,][]{Iba01,DB02,Nav04,BJ05}.

While both major and minor mergers are expected to produce some degree
of morphological evolution, it is still unclear whether all mergers
will lead to a merger--driven starburst and therefore drive
spectrophotometric evolution. There is mounting evidence that near
equal mass (major) mergers trigger the most vigorous
star-forming galaxies in the local Universe, the ultraluminous
infrared galaxies (ULIRGs) with bolometric luminosities greater
$10^{12}$~\lsun\ \citep[see, e.g.,][]{SM96,Bor00}.  More generally,
large statistical samples consistently indicate a clear
anticorrelation between projected galaxy pair separation and star
formation indicators in the optical \citep{Bar00,NCA04,Lam03,SA06,
Bar07,Ell07} and the infrared \citep{Gel06,SS07}.  Moreover, numerical
simulations naturally explain these observations as centrally
concentrated starbursts triggered by tidal forces that attend the
galaxy interaction \citep[see,
e.g.,][]{BH91,MH94majm,MH96,Sp00,Kap05,Cox06,dM07}.

In contrast, relatively little is known about the resulting starbursts during
interactions between galaxies of unequal mass.  Observationally, statistical studies
are hampered by magnitude limits and confusion with background sources, but
preliminarily indicate that enhanced star formation (as measured by H$\alpha$) is
strongly dependent upon the relative mass (or magnitude) of the interacting galaxies.
Initial results using the CfA2 Redshift Survey indicated no correlation between 
galaxy pair separation and equivalent width of H$\alpha$ \citep{WGB06}, suggesting
that minor mergers do not lead to merger--driven star formation.  More recent studies
using the Sloan Digital Sky Survey \citep[SDSS,][]{Ell07,WG07} find that minor mergers
indeed lead to enhanced star formation, however the data also suggest that the enhancement
is primarily observed in the satellite galaxy while the larger primary galaxy has star
forming properties similar to non--interacting field galaxies \citep{WG07}.
More locally, detailed observations of individual galaxies in the nearby Universe suggest
that minor mergers may be responsible for some degree of star formation enhancement,
e.g., NGC 278 \citep{Kna04}, NGC 3310 \citep{Smi96},  NGC 4064 and NGC 4424 \citep{CKH06},
NGC 7742 \citep{Maz06}, and the M81 group \citep{Wal02,FS03}.

While numerical simulations have been used extensively in the study of
star formation in major mergers, there exist only a handful of
previous simulations that have specifically quantified the star
formation history of unequal mass galaxy mergers.  These works have
established that tidal perturbations can induce bar formation, inflows
of gas, and subsequent bursts of star formation when the primary disks
are susceptible to instabilities, such as when they do not contain a
stellar bulge \citep{H89,MH94minm,HM95}.  However, because these
studies performed only a limited number of simulations, and did not
include some important physical processes, our understanding of the
relationship between mergers of unequal mass galaxies and the
resulting star formation is still highly incomplete.

In this paper we fill in this gap by describing a large set of numerical simulations
of unequal mass disk-galaxy mergers.  In particular, we seek to extend and improve
the pioneering work of \citet{MH94minm} and provide a more complete picture of 
merger--driven star formation, including its dependence upon the merging orbit and
the structure of the primary galaxy.  We will also take advantage of recent improvements
in the treatment of supernovae feedback \citep{Sp00,SH03,Cox06} and smoothed--particle
hydrodynamics (SPH) methodology \citep{SHEnt} which allow us to follow the evolution of
galaxies with larger gas fraction than was possible previously.

The results of our simulations are relevant for a number of related studies.  Many
models of galaxy formation often employ merger--driven star formation as a necessary
ingredient to produce bright galaxies throughout cosmic time \citep{Gui98,SPF,Bau05}.
Our results will serve as a useful input to such models.  In addition, our
models will be a valuable tool for interpreting observations.  By comparison to the
observed star formation in pairs of galaxies \citep{Bar00,WGB06,WG07,Lin07}, we can constrain
our models and infer the contribution of mergers to the universal star formation rate.
The numerical models have also been used to generate ``simulated observations'' of the
merging galaxies through radiative transfer modeling \citep{Patrik,Jon06,JonSun},
to study dust attenuation in spiral galaxies \citep{Roc07}, to understand the size,
shape, and scaling relations of merger remnants \citep{DC06,Nov06,Cov07}, and to calibrate
non-parametric indicators of galaxy morphology (Lotz et al., in preparation), which
can then be used to quantify the merger rate over cosmic time.

We organize this paper as follows:
Our numerical techniques are summarized in \S\ref{sec:sims} and the galaxy models 
used for our galaxy interactions are described in \S\ref{sec:ics}.
A general description of the mergers is provided in \S\ref{sec:mergers}, including
a discussion about the possible methods to quantify the merger-driven star
formation.   Some additional models are explored in \S\ref{sec:omods}.  Finally,
we discuss our results and conclude in \S\ref{sec:disc}.

\section{Numerical Simulations}
\label{sec:sims}

All of the numerical simulations performed in this work use the N-Body/SPH code GADGET
\citep{SpGad}.  In general, our methodology is similar to \citet{Thesis} and 
\citet{Cox06}, and thus, in this section we will only review selected aspects
of our techniques and methodology that are relevant to this work.  We use the
first version of GADGET, however the smoothed particle hydrodynamics (SPH)
modules are upgraded (with the aid of V. Springel) to use the ``conservative
 entropy'' version that is detailed in \citet{SHEnt}.  
The radiative cooling rate $\Lambda_{\rm net}(\rho,u)$ is computed
for a primordial plasma as described in \citet{Kz96}.  Stars are formed at a 
rate determined by the local SPH density normalized to match observed star
formation rates \citep{Kenn98}.  Furthermore, we employ a threshold density 
$\rho_{\rm th}$, below which stars do not form.  Finally, star formation occurs
within individual SPH particles and each particle can stochastically spawn two
stellar particles.

One of the largest uncertainties associated with current numerical simulations including star
formation is the implementation of supernova feedback.  Because of the limited
resolution achievable in state-of-the-art numerical simulations, most work performed
to date adopts a ``sub--grid'' approach where the physical processes associated with
feedback are included in a simple, yet flexible manner with free parameters that
can be tuned to match observations.  Unfortunately, there are often a range of
acceptable free parameters which produce distinct star formation histories
or remnant morphologies \citep[see, e.g.,][]{TC00,Kay02,SdMH05,Cox06}.  For many of
the simulations performed in this work, we have run two different feedback models,
either the $n0med$ or $n2med$ models as introduced in \citet{Cox06}.  Both models are termed
``medium'' because they dissipate feedback energy on a 8~Myr timescale.  The $n2med$ 
model treats star-forming gas with a stiff equation of state while $n0med$ assumes
this gas is isothermal with an effective temperature $\sim10^5$~K.  We will compare these two
feedback models in \S\ref{sssec:sffbps} in order to determine how these assumptions
influence the resulting star-formation history.

All simulations presented throughout this work use a gravitational
softening length of 0.1~kpc for all baryonic particles, and 0.4~kpc
for dark matter particles.  The SPH smoothing length is required to be
greater than half the gravitational softening length, or $>50$~pc.  We
have performed a few tests with both smaller and larger softening
lengths and found little or no differences in the star-formation
history, suggesting that we have achieved numerical convergence
\citep[see][for similar tests]{Cox06}.

\section{Galaxy Models}
\label{sec:ics}

Because our primary motivation is to investigate collisions between unequal mass 
galaxies, we wish to construct galaxies with a range of different masses.  
Furthermore, because the properties of galactic disks depend on mass, we also need
to design galaxy models that are physically plausible.  To begin, we select the
largest galaxy model to have a total stellar mass of 
M$_{\star}\sim$~5.0$\times10^{10}$~\msun, slightly above the transitional mass of
3.0$\times10^{10}$~\msun~found to divide large galaxies with old stellar 
populations from smaller star forming ones \citep{Kau03}.  Three smaller galaxy
models are included that have M$_{\star}$ equal to 1.5 $\times10^{10}$~\msun, 0.5 
$\times10^{10}$ \msun~and 0.1 $\times10^{10}$~\msun, making the total stellar
mass span a range of 50.  We label the largest of these models G3, and the rest
G2, G1, and, finally, the smallest G0.

\begin{table*}
\begin{center}
\caption{Properties of galaxy models used in this study.  The compound
galaxy has a total mass M$_{\rm vir}$, and is composed of dark matter
and a stellar disk, stellar bulge, and gaseous disk.  The total mass
of each baryonic component and their mass fractions are provided.  The
dark halo has an NFW profile with concentration $c$ and spin parameter
$\lambda$.  The stellar disk has a total mass M$_{\rm disk}$ which is
a fraction m$_d$ of the total mass and a fraction j$_d$ of the total
angular momentum.  The disk has an exponential profile of scale radius
R$_d$ and scale height z$_0$.  Each galaxy model contains a spherical
bulge of mass M$_{\rm bulge}$ and mass fraction m$_b$, with an
exponential profile of scale length R$_b$ and zero angular momentum.
The gas mass is M$_{\rm gas}$, a fraction m$_g$ of the total mass, and
j$_g$ is the angular momentum fraction.  The gas disk, like its
stellar companion, is assumed to have an exponential surface density
with a scale radius $\alpha$ times R$_d$.  We define the gas fraction
$f$ as the fraction of the {\it total} disk mass which is gaseous.
Finally, each system is represented by N particles, which represent
the dark matter, stellar disk, bulge and gaseous galaxy.  }
\begin{tabular}{lcccc}
\hline 
 & G3 & G2 & G1 & G0 \\
\hline
\hline
M$_\star$ ($10^{10}$ \msun)                    &   5.0   &   1.5   &   0.5   &   0.1   \\
R$_{50}$ (kpc)                                 &   3.86  &   2.88  &   2.35  &   1.83  \\
\hline
Total Mass, M$_{\rm vir}$ ($10^{10}$ \msun)    & 116.0   &  51.0   &  20.0   &   5.0   \\
Concentration, c=R$_{\rm vir}/r_s$             &   6     &   9     &  12     &  14     \\
Spin Parameter, $\lambda$                      &   0.05  &   0.05  &   0.05  &   0.05  \\
\hline
Disk Mass, M$_{\rm disk}$ ($10^{10}$ \msun)    &   4.11  &   1.35  &   0.47  &   0.098 \\
Disk Mass Fraction, m$_d$                      &   0.035 &   0.026 &   0.024 &   0.019 \\
Disk Scale Length, R$_d$ (kpc)                 &   2.85  &   1.91  &   1.48  &   1.12  \\
Disk Scale Height, z$_0$ (kpc)                 &   0.40  &   0.38  &   0.30  &   0.22  \\
Disk Spin Fraction, j$_d$                      &   0.015 &   0.010 &   0.010 &   0.010 \\
\hline
Bulge Mass, M$_{\rm bulge}$ ($10^{9}$ \msun)   &   8.9   &   1.5   &   0.3   &   0.02  \\
Bulge-to-disk ratio, B/D                       &   0.22  &   0.11  &   0.06  &   0.02  \\
Bulge Mass Fraction, m$_b$                     &   0.008 &   0.003 &   0.002 &$<$0.001 \\
Bulge Radial Scale Length, R$_b$ (kpc)         &   0.37  &   0.26  &   0.20  &   0.15  \\
\hline
Gas Mass, M$_{\rm gas}$ ($10^{10}$ \msun)      &   1.22  &   0.48  &   0.20  &   0.06  \\
Gas Mass Fraction, m$_g$                       &   0.011 &   0.009 &   0.010 &   0.012 \\
Gas Fraction, f                                &   0.196 &   0.242 &   0.286 &   0.375 \\
Gas Scale Multiplier, $\alpha$                 &   3.0   &   3.0   &   3.0   &   3.0   \\
Gas Spin Fraction, j$_g$                       &   0.012 &   0.010 &   0.013 &   0.019 \\
\hline
N                                              & 240,000  & 150,000 & 95,000  & 51,000 \\
N$_{\rm dm}$                                   & 120,000  & 80,000  & 50,000  & 30,000 \\
N$_{\rm gas}$                                  &  50,000  & 30,000  & 20,000  & 10,000 \\
N$_{\rm disk}$                                 &  50,000  & 30,000  & 20,000  & 10,000 \\
N$_{\rm bulge}$                                &  10,000  & 10,000  &  5,000  & 1,000  \\
\hline 
\end{tabular}
\label{tab:ics}
\end{center}
\end{table*}

Once the stellar masses are fixed, we next determine the stellar size using the 
late-type galaxy size-mass relationship found by \citet[eq. 18]{Shen03}, from
the analysis of 140,000 galaxies in the Sloan Digital Sky Survey.  This relation
fixes the half-light radius R$_{50}$ for each model, which we assume is equivalent
to the stellar half-mass radius.

Because each model is assumed to be a late-type galaxy, they consist of a stellar
and gaseous exponential disk, a stellar \citet{H90} bulge, all embedded in a dark
matter halo.  A general description of the methods employed to construct model
disk galaxies, including the assumed profiles, can be found in \citet{Cox06},
however, those models are not identical to the ones used here.  The model
parameters used for this work, which are listed in Table~\ref{tab:ics}, are selected
such that each model galaxy is statistically average, as opposed to \citet{Cox06} who
specifically modeled a gas--rich Sbc galaxy.

To ensure that the models are statistically average, the
mass of the stellar disk and bulge are constrained by the half-mass radius R$_{50}$
and the observed bulge-to-disk ratio of local late-type galaxies \citep{dJ96}.  To
quantify the latter constraint, we assume a K-band mass-to-light ratio of 0.7 and
0.5 for the bulge and disk, respectively, and fit a line to the \citet{dJ96} data,
finding
\be
\log ({\rm M}_{\rm bulge}) = 1.6~\log ({\rm M}_{\rm disk}) - 1.03,
\label{eq:fitdJ}
\ee
where the masses are in units of $10^{10}$~\msun.  Using this formula, along with
the fixed total stellar mass for each model, both the stellar bulge and stellar disk
masses are determined.  To determine the size of these components (the exponential
disk scale radius R$_d$ and the bulge scale radius R$_b$) an iterative approach is
used where the first step is to guess an initial disk scale radius R$_d$.  The bulge
scale radius R$_b$ is then fixed by the empirical correlation between disk and bulge
sizes found by \citet[][see his Eq.~5]{dJ96}.  This initial guess for the disk and
bulge mass distributions is then compared to the desired half-mass radius R$_{50}$
and corrected until the two are within 1\% of each other.

In addition to the stellar component, each galaxy model contains gas,
the fuel for star formation.  Motivated by observations of local
late-type galaxies \citep[e.g.,][]{BvW94}, gas is distributed in an
extended exponential disk with a scale radius proportional to that of
the stellar disk, $r_{\rm gas} = \alpha r_{\rm star}$. In our default
model we assume $\alpha=3$.  To determine the mass of gas, we use the
observed trend that systems with lower total/stellar mass have a
higher gas fraction \citep{Bell03,Geh06}.  Using data from
\citet[][kindly provided in electronic form by E. Bell]{Bell03}, we
parameterize this relationship as \be \log ({\rm M}_{\rm gas}) =
0.78~\log ({\rm M}_\star) + 1.74,
\label{eq:fitbell}
\ee
where both masses are in units of $10^{10}$~\msun.  Because this relation is derived
from the mean gas content as a function of total baryonic mass, the model disks will
contain only moderate quantities of gas, and are therefore unlikely to produce large
bursts of star formation.  In \S\ref{ssec:gasf} we will perform a small number of 
additional mergers using galaxies with higher gas fractions.

The final component of the galactic system is the massive dark-matter
halo.  Each model contains a dark halo with a \citet[][NFW]{NFW96}
profile whose properties are selected so that the rotation curve
satisfies the baryonic Tully-Fisher relation \citep{BdJ01,Bell03}.
Because most of the models require very little non-baryonic mass in
order to satisfy the Tully-Fisher relation, we do not include
adiabatic contraction.  Still, the resultant halo concentrations $c$
are below the mean found in cosmological N-body simulations
\citep{Bul01}, a tension that has been emphasized previously
\citep[see, e.g.,][]{AB02}.  We note that (motivated by observations)
the galaxy models have systematically higher baryon fractions in
higher mass galaxies, resulting in a total mass ratio 23 between G3
and G0 as opposed to a ratio of 50 in stellar mass.

The construction of these galaxy models is based upon \citet{H93} and its more recent
incarnation to include a NFW halo \citep{SW99,Sp00}.  We note that because the spatial
distribution of baryons is fixed by the galaxy model, the halo spin does not determine
the size of the gaseous and stellar disk, as in many popular models for disk formation
\citep[see, e.g.,][]{MMW98}.  However, we feel that our procedure is adequate since it
produces stable disks, provided that the velocity dispersion of the disks is chosen
such that the Toomre stability parameter $Q$ is greater than 1 at all radii and the
feedback model can provide adequate pressure support to the gas.

\subsection{Evolution of Isolated Systems}
\label{ssec:isosf}

\begin{figure*}
\begin{center}
\resizebox{4.1cm}{!}{\includegraphics{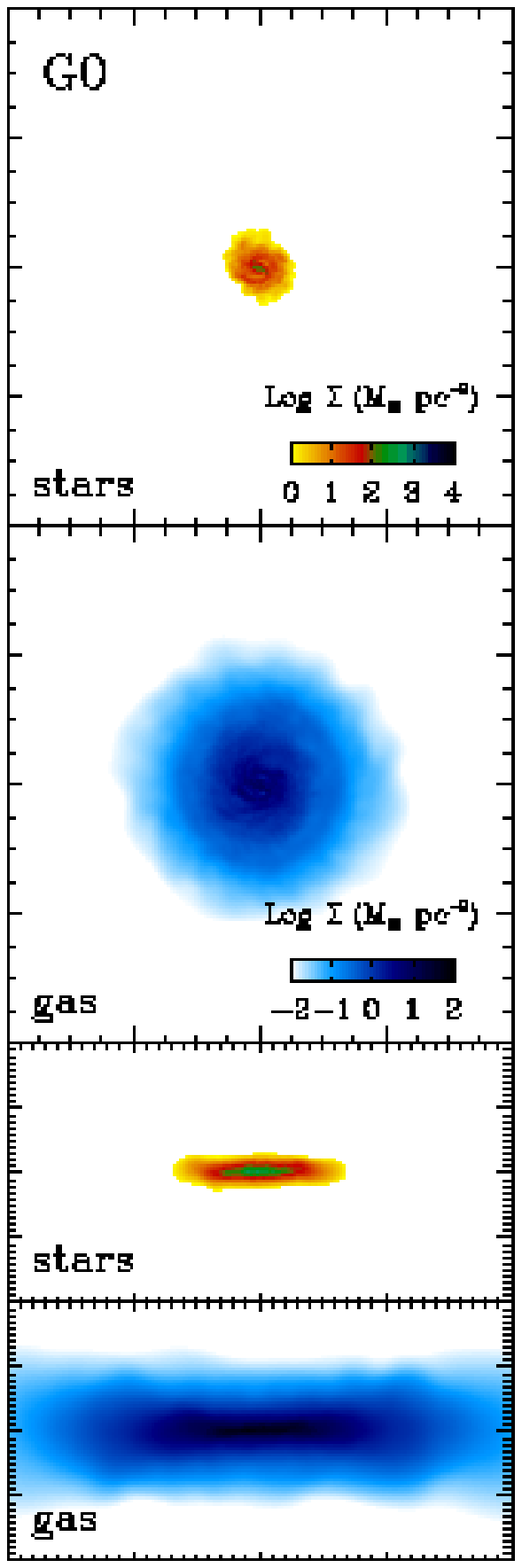}}%
\resizebox{4.1cm}{!}{\includegraphics{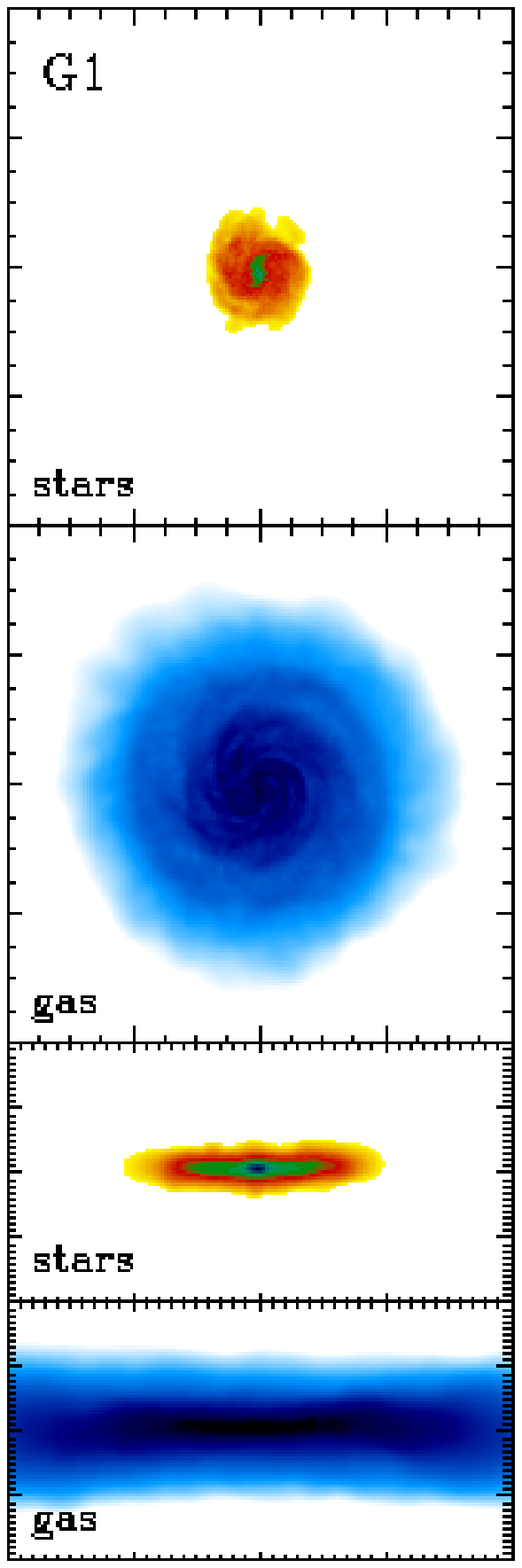}}%
\resizebox{4.1cm}{!}{\includegraphics{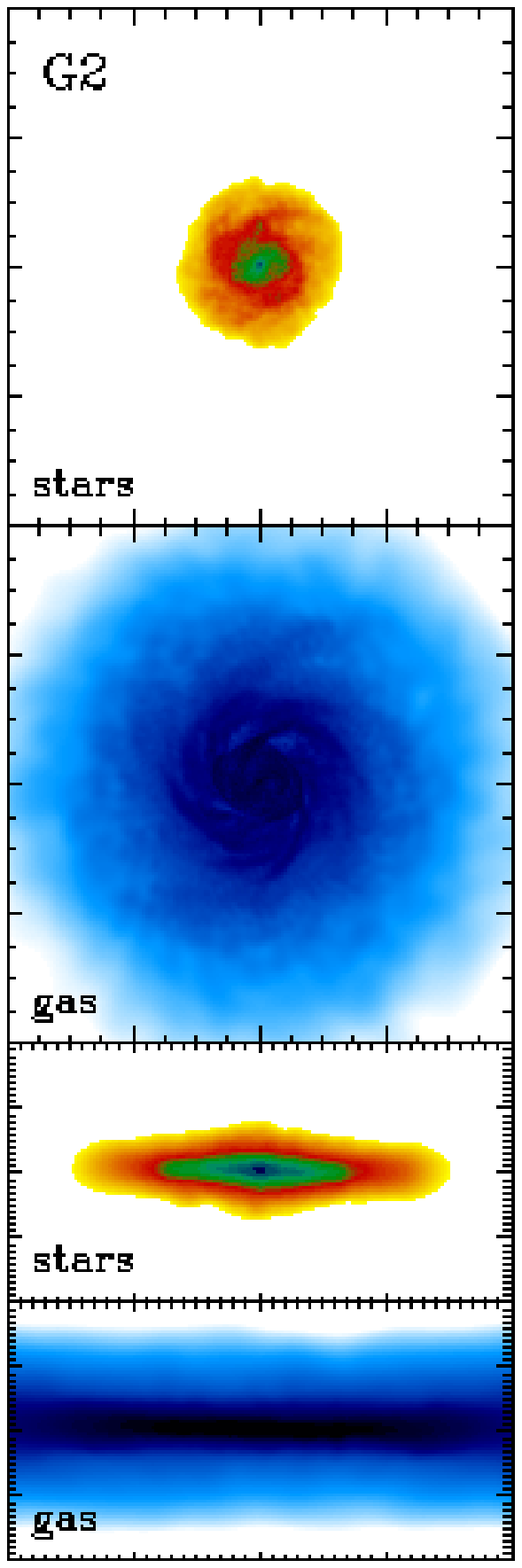}}%
\resizebox{4.1cm}{!}{\includegraphics{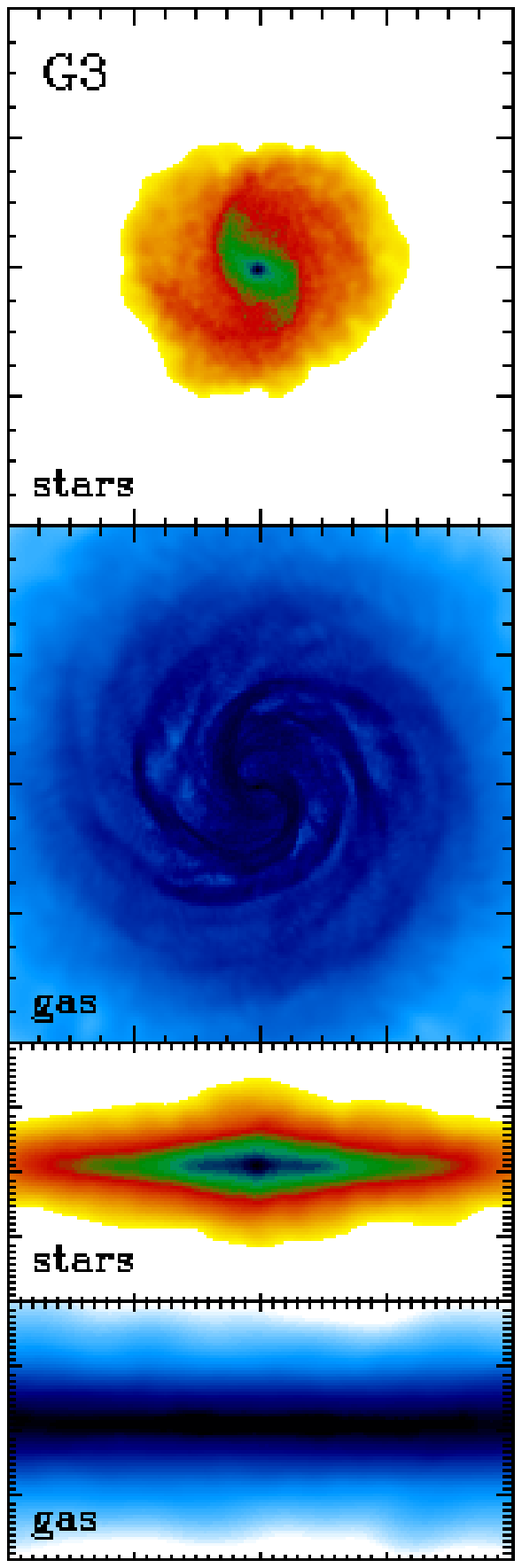}}\\
\caption{Projected mass density for each galaxy model, G0 (left column), G1, G2, and 
G3 (right column) when simulated in isolation for $\sim1$~Gyr, which is greater than
5 orbital periods at the half--mass radius in all models.  Each row shows the 
projected stellar or gaseous mass as specified in the lower-left of each panel.  The
top two rows show the galaxies face--on, while the bottom two rows show the edge--on
view.  Each panel measures 80~kpc on a side.  A color bar in the upper--left panel
indicates the stellar surface density scale that is used for all panels, and the
color bar in the one below indicates the gaseous surface density scale that is used
in all panels.
\label{fig:isomorph}}
\end{center}
\end{figure*}

The stability of our model galaxies is demonstrated in Figure~\ref{fig:isomorph},
which shows the projected gas and stellar surface densities after being evolved
in isolation for $\sim1$~Gyr (greater than 5 orbital periods in all models).  While
the mass distributions of each component are stable during the entire period that
they are simulated, the noisy potential resulting from the finite particle number
seeds transient instabilities \citep[see][]{H93}.  These instabilities are most prominent
in the G3 and G2 galaxies, which have larger disk surface densities than the
less massive galaxy models.

\begin{figure}
\begin{center}
\resizebox{8.0cm}{!}{\includegraphics{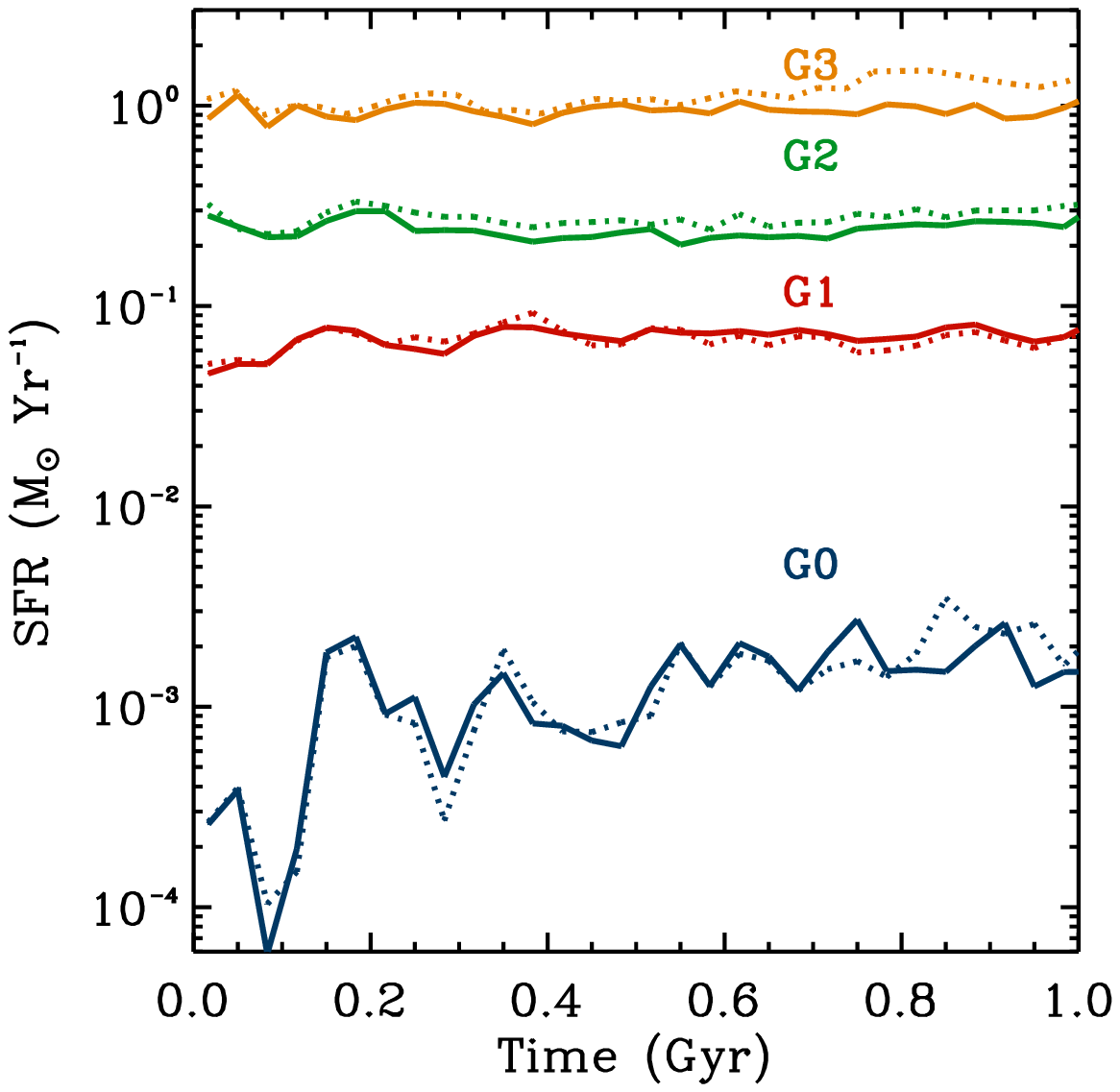}}\\
\caption{Star formation for our isolated galaxy models
G3, G2, G1 and G0.  Each model has been run with two separate
feedback models and both star-formation histories are shown,
$n0med$ with a dotted line and $n2med$ with a solid line.
\label{fig:sfriso}} 
\end{center}
\end{figure}

The stability of the galaxy models is also reflected by a constant star-formation
rate (SFR), regardless of the feedback model employed, as shown in 
Figure~\ref{fig:sfriso}.  The average SFR is 0.95, 0.25, 0.06
and 0.001 \msunyr\ for models G3, G2, G1, and G0, respectively, during the 1~Gyr
period shown in Figure~\ref{fig:sfriso}.  The average SFR scales as
M$_{\rm gas}^{3/2}$, as is expected from our Schmidt-type star-formation law
\citep[see][]{Cox06}.  The SFR of the smaller galaxies, G1 and G0, display increased
fluctuations owing to the more significant impact of feedback, decreasing resolution, and the
fixed star-formation threshold density $\rho_{\rm th}$.  The SFRs shown in
Figure~\ref{fig:sfriso} are constant for $\sim2$~Gyr.  After this, galaxies G3 and G2
begin to consume a significant fraction of their high-density ($>\rho_{\rm th}$) gas
and their SFR eventually drops to approximately one--tenth its initial value at 6~Gyr.  Another 
consequence of the adopted star-formation law is that the star-formation timescale
scales with gas density, viz., $\tau_{\rm SF}\approx$ M$_{\rm gas}$/SFR~$\sim\rho^{-1/2}$, in other
words, more massive galaxies, with higher gas densities, have shorter gas consumption
timescales.  As a result, the more massive galaxy models consume a larger fraction of
their initial gas mass than the smaller ones.
During the entire 6~Gyr evolution, model G3 converts 26\% of its original gas
to stars, G2 converts 22\%, G1 converts 11\%, and G0 converts 3\%.

\section{Galaxy Mergers}
\label{sec:mergers}

The previous section defined four galaxy models G3-G0 and established the
stability of these models by simulating them in isolation.  In this section
we discuss the outcome of binary mergers between these models on a single
merging orbit.  Performing a single merger between all combinations of the
four models yields four equal-mass major mergers and six unequal--mass
mergers.  In the following section (\S\ref{sec:omods}), we discuss a 
number of additional merger simulations to determine systematic 
dependencies.

\subsection{Fiducial Encounters}
\label{ssec:orbits}

\begin{table}
\begin{center}
\caption{
Total, stellar, and baryonic mass ratios between our four
galaxy models G3, G2, G1, and G0.
}
\begin{tabular}{llccc}
\hline
Primary & Satellite & Total & Stellar & Baryonic \\
\hline
\hline
G3 & G3 &   1:1 &   1:1 &   1:1 \\
G3 & G2 & 2.3:1 & 3.3:1 & 3.1:1 \\
G3 & G1 & 5.8:1 & 10.0:1 & 8.9:1 \\
G3 & G0 & 22.7:1 & 50.0:1 & 38.9:1 \\
\hline
G2 & G2 &  1:1 & 1:1 & 1:1  \\
G2 & G1 & 2.6:1 & 3.0:1 & 2.8:1  \\
G2 & G0 & 10.0:1 & 15.0:1 & 12.4:1  \\
\hline
G1 & G1 & 1:1 & 1:1 & 1:1  \\
G1 & G0 & 3.9:1 & 5.0:1 & 4.4:1  \\
\hline
G0 & G0 & 1:1 & 1:1 & 1:1  \\
\hline
\end{tabular}
\label{tab:massratios}
\end{center}
\end{table}

For reference, we list the merger mass ratios in Table~\ref{tab:massratios}.  Mass
ratios are given for the total, stellar and baryonic masses.  In what follows
we will label a simulated mergers by its two merging constituents.  For example,
G3G1 is a merger between our largest galaxy G3 and the second--smallest G1.  The
total mass ratio is 5.8 to 1, i.e., model G1 is 5.8 times smaller than G3.  When
referring to the interacting galaxies, the less massive galaxy will be termed the
satellite and the more massive galaxy the primary.  We note that some galaxy
models (G2, G1, and G0) can be a primary in one interaction and a satellite
in another.

All galaxy interactions are initialized with a nearly unbound elliptical orbit
with eccentricity $\epsilon=$0.95.  The initial separation is fixed to be slightly 
less than the virial radius of the primary and is 250, 100, 80, and 50~kpc for G3,
G2, G1, and G0 models, respectively.  The pericentric distance is set to be 13.6,
3.8, 2.96, and 2.24 when the primary is G3, G2, G1, and G0, respectively.
The resulting interactions are fast and nearly radial, consistent with orbits
found for dark matter halos in cosmological simulations \citep{Vit02,KB04,Ben05,Zent05}.
The only exception to the above orbits is the merger G3G0, where the total mass
ratio is 22.7 to 1, the largest mass ratio we simulate.  In this case, the fiducial
interaction was still not fully merged after the system was evolved for
$\sim12$~Gyr.  In order to better compare this interaction with the other, more
rapid, mergers, we multiplied the initial velocity of the satellite by 0.2 so that
the merger occurred during the 6~Gyrs we followed each interaction.

Finally, the fiducial series of orbits are all prograde, i.e., the angular momentum
of the orbit is (nearly) aligned with the spin of the primary.
A slight tilt of 30$^\circ$ is introduced such that the interaction is not entirely
co-planar.  In section~\ref{ssec:morb} we resimulate the G3G2 and G3G1 mergers
with a variety of orbits to determine the dependence of merger--driven star
formation on orbital angular momentum (R$_{\rm peri}$) and merger alignment.

\subsection{Merger Evolution}
\label{ssec:mm}


\begin{figure*}
\resizebox{16.0cm}{!}{\includegraphics{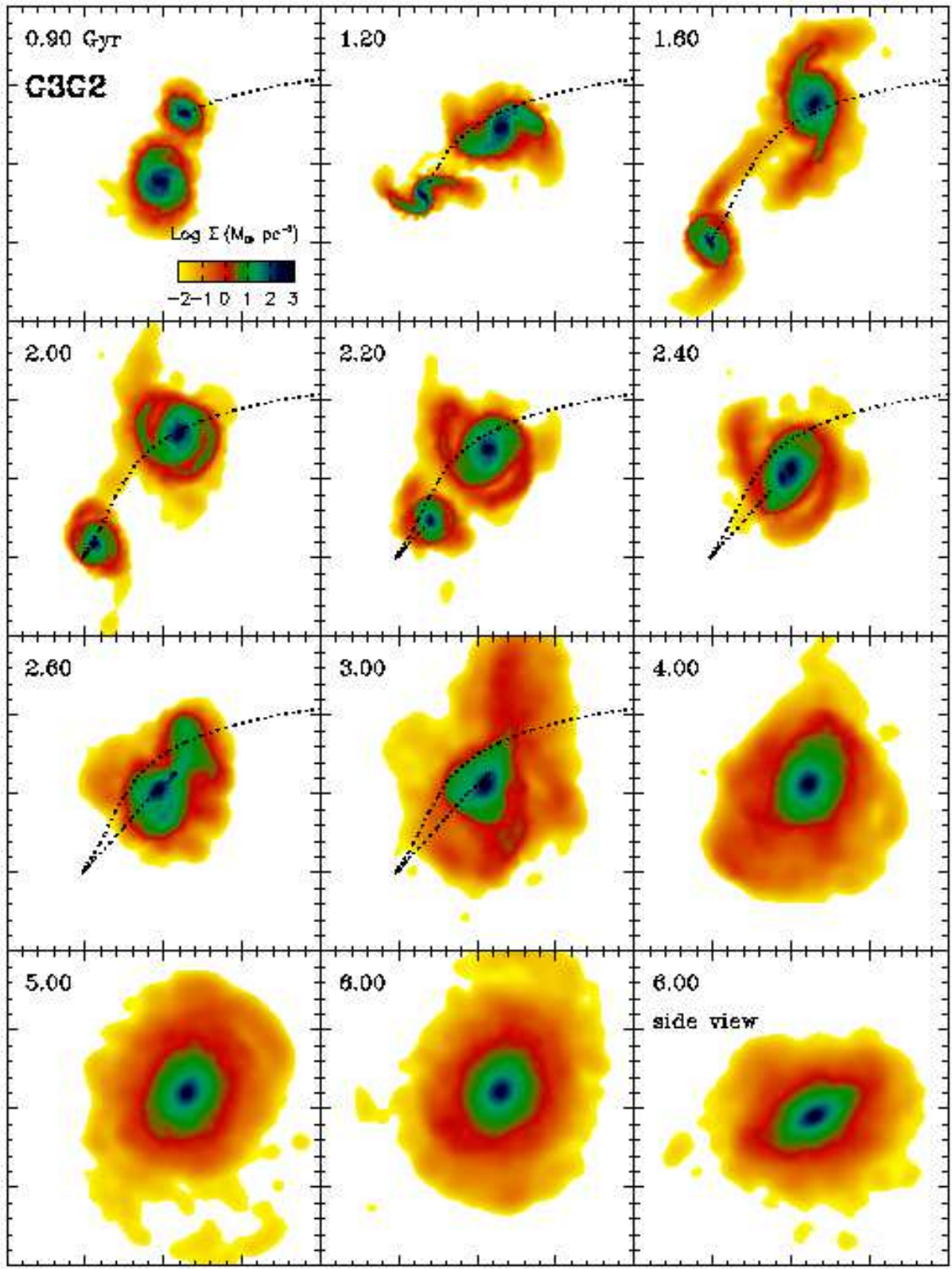}}\\
\caption{Projected stellar mass density during the G3G2 merger simulation
as viewed in the orbital plane.  Each panel measures 200~kpc on a side
and the time, in Gyr, is displayed in the upper left of each panel.  The
orbit of the satellite galaxy G2 is denoted by a dotted line until it has
completely merged with the primary galaxy.  The bottom--right panel shows
a side view of the final merger remnant, and clearly indicates the initial
30$^\circ$ inclination of the progenitor disk.  The top--left panel
indicates the color scale used in all panels.
\label{fig:G3G2stellarmorph}}
\end{figure*}

\begin{figure*}
\resizebox{16.0cm}{!}{\includegraphics{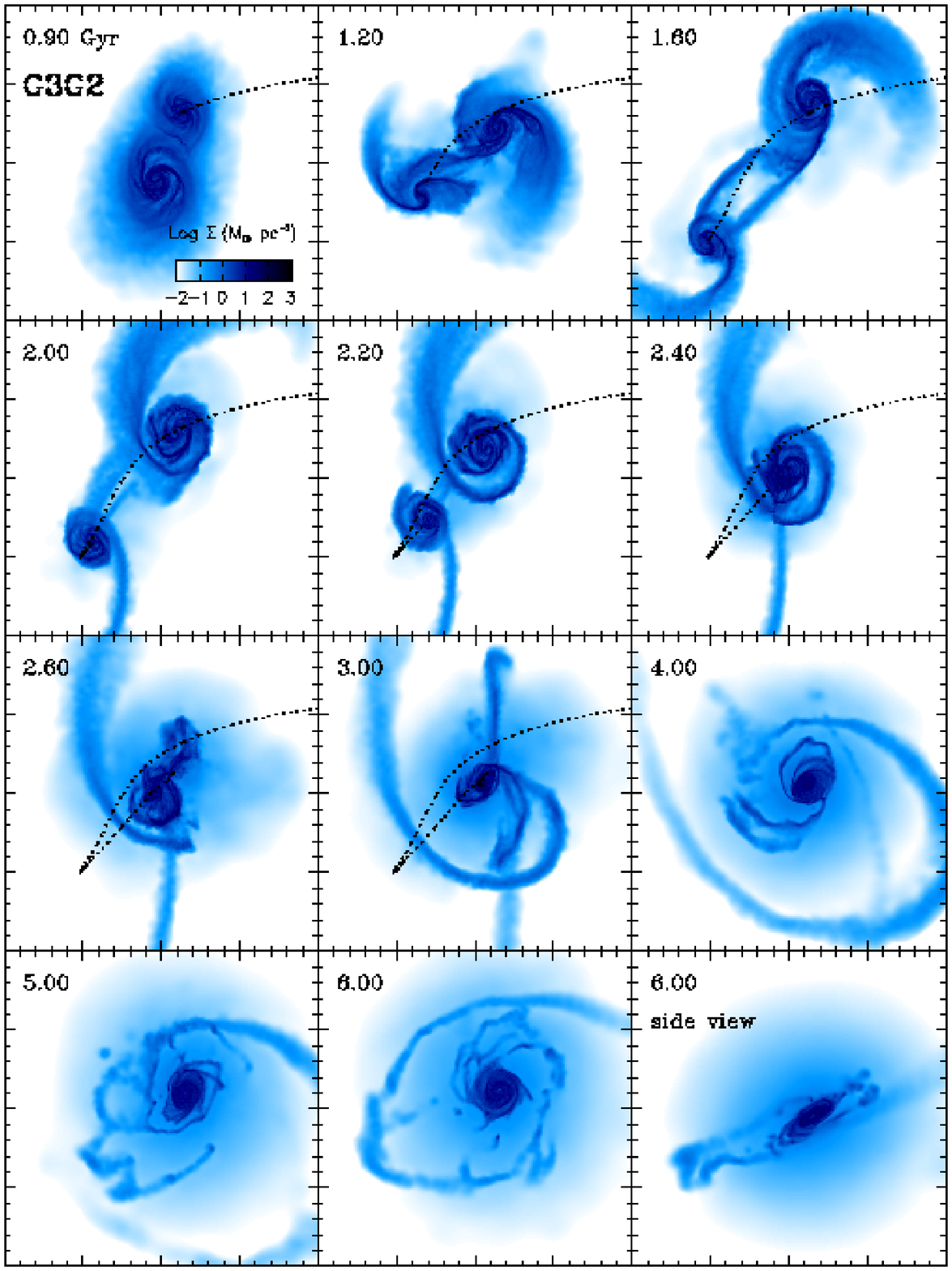}}\\
\caption{Similar to Figure~\ref{fig:G3G2stellarmorph}, but for the
gaseous component.
\label{fig:G3G2gasmorph}}
\end{figure*}

\begin{figure*}
\resizebox{16.0cm}{!}{\includegraphics{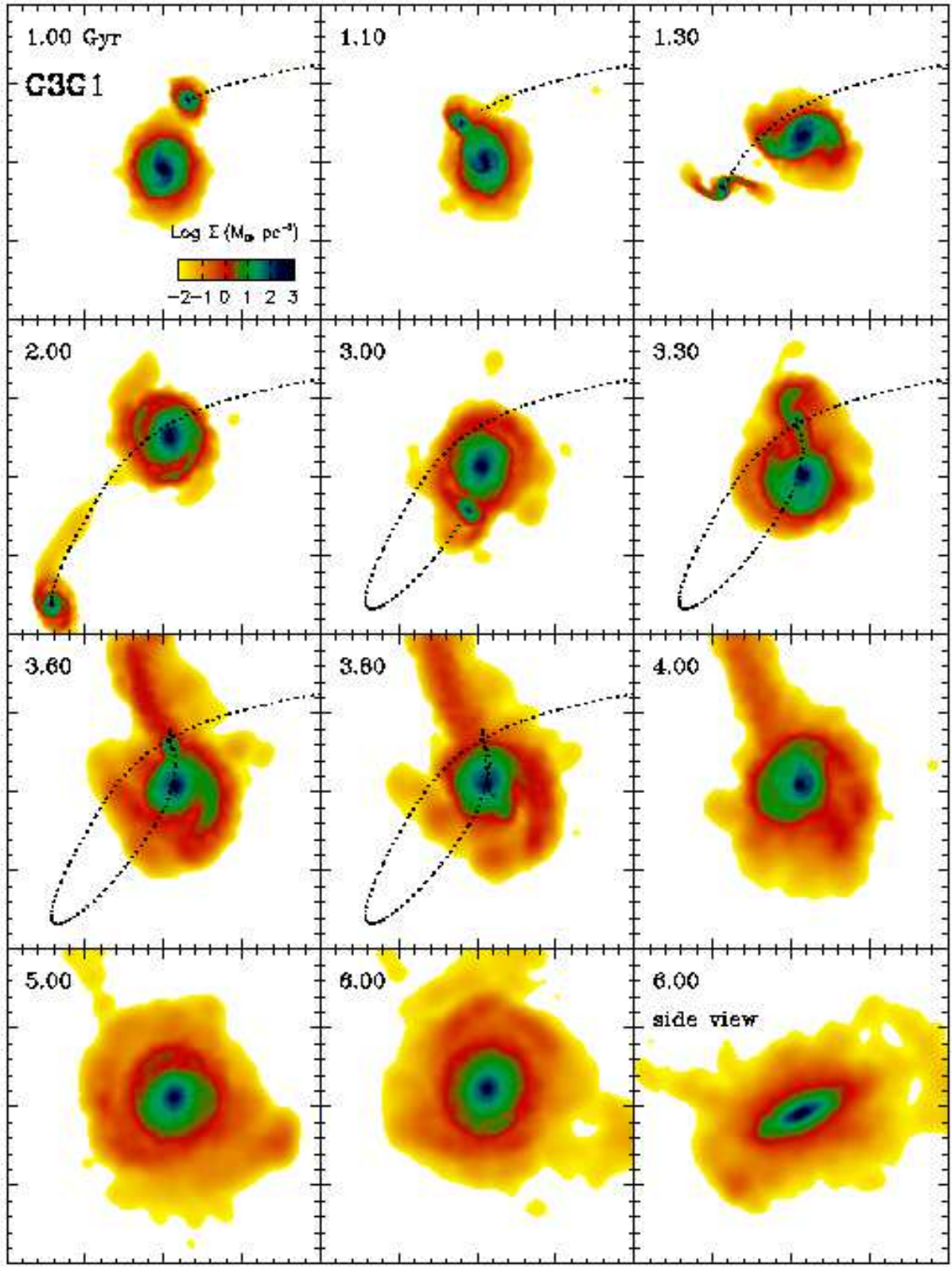}}\\
\caption{Similar to Figure~\ref{fig:G3G2stellarmorph}, but the
G3G1 merger is now shown.
\label{fig:G3G1stellarmorph}}
\end{figure*}

\begin{figure*}
\resizebox{16.0cm}{!}{\includegraphics{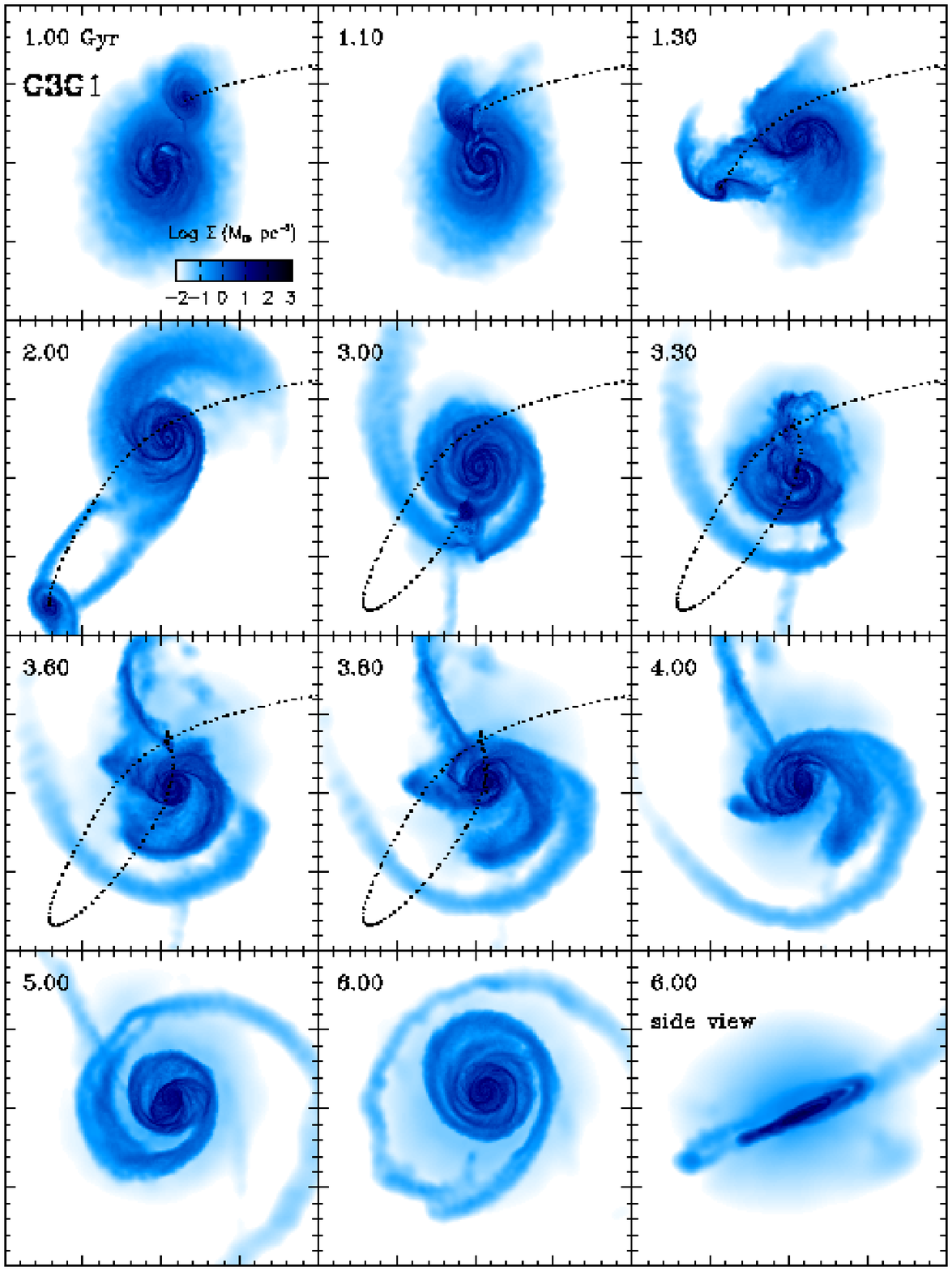}}\\
\caption{Similar to Figure~\ref{fig:G3G1gasmorph}, but for the
gaseous component.
\label{fig:G3G1gasmorph}}
\end{figure*}

The evolution of a typical merger event is shown in Figures~\ref{fig:G3G2stellarmorph}
and \ref{fig:G3G2gasmorph} for the G3G2 merger, and Figures~\ref{fig:G3G1stellarmorph}
and \ref{fig:G3G1gasmorph} for the G3G1 merger.  In both cases, the projected
stellar surface density is first shown, followed by the projected gaseous surface
density.  The images shown in these figures are typical of the remaining mergers
that are not shown, and are also consistent with what has been discussed in 
studies of equal mass mergers \citep[e.g.,][]{BH91,MH96,Sp00,Cox06}, and unequal mass
mergers \citep[e.g.,][]{H89,HM95}.  For completeness, the following provides a brief
outline of the merger process, and the resulting star formation.

In both of the mergers shown the satellite galaxy (G2 and G1) first makes a fast,
direct approach toward the primary galaxy (G3).  This close passage tidally distorts
the disks and generates symmetric tails in both.  Owing to the energetic orbit, the
satellite separates again for several orbital periods ($\geq$1~Gyr) before returning
for its second encounter with the disk of the primary.  A generic
feature of all the interactions simulated here is the efficient loss of angular
momentum by the satellite.  In particular, even for large mass ratio encounters the
orbits become almost entirely radial after the first or second close passage.

The loss of orbital angular momentum and eventual radial nature of the final merger
limits the coupling between orbital and spin angular momentum, and therefore the
tidal response.  Most of the induced star formation during the final coalescence is
a result of
the collisional nature of the gas and the rapid fluctuations in the potential as
the satellite oscillates through the center and the system finally relaxes.  The
collisional nature of the gas also produces several other features distinct from
the stellar disk.  During the first close passage, the satellite creates a
snow--plow effect, as it clears out gas from the primary's extended
gas disk.  This effect continues during each close passage until eventually the
majority of the gas in the satellite has been stripped off, leaving a predominantly
stellar satellite.  However, by this point the orbit has decreased such that it
is difficult to disentangle the two galaxies.

While it is beyond the scope of the current paper to provide a detailed analysis of
the structure of the merger remnants, there are several qualitative features present
in Figures~\ref{fig:G3G2stellarmorph}--\ref{fig:G3G1gasmorph} that motivate future
inquiry.  First, it is evident that the stellar disks have been thickened and
dynamically heated during the accretion event, consistent with a number of previous
numerical studies \citep[e.g.,][]{QG86,Qui93,TO92,WMH96,VW99,Ben04,BCJ04,BJC05}.  The side
views demonstrate that the G3G2 interaction has had a more disruptive effect on the
primary's stellar and gaseous disk than the G3G1 interaction,
an unsurprising result given the smaller mass ratio and hence larger tidal disturbance.
Furthermore, the gaseous disks are also affected.  It is particularly interesting that
the G3G1 remnant contains a relatively large, thin gaseous disk, indicating that during
an unequal mass interaction the gaseous disk can dissipate its energy and maintain its
structure while the stellar disk cannot.  The
second qualitatively interesting feature is the efficient deposition of the satellite
galaxy in the center of the merger remnant.  Such a process may be a prominent mechanism
to build galactic bulges \citep[see also][]{ABP01,EM06}.

\subsection{Star Formation}
\label{ssec:sfr}

\begin{figure*}
\begin{center}
\resizebox{8.0cm}{!}{\includegraphics{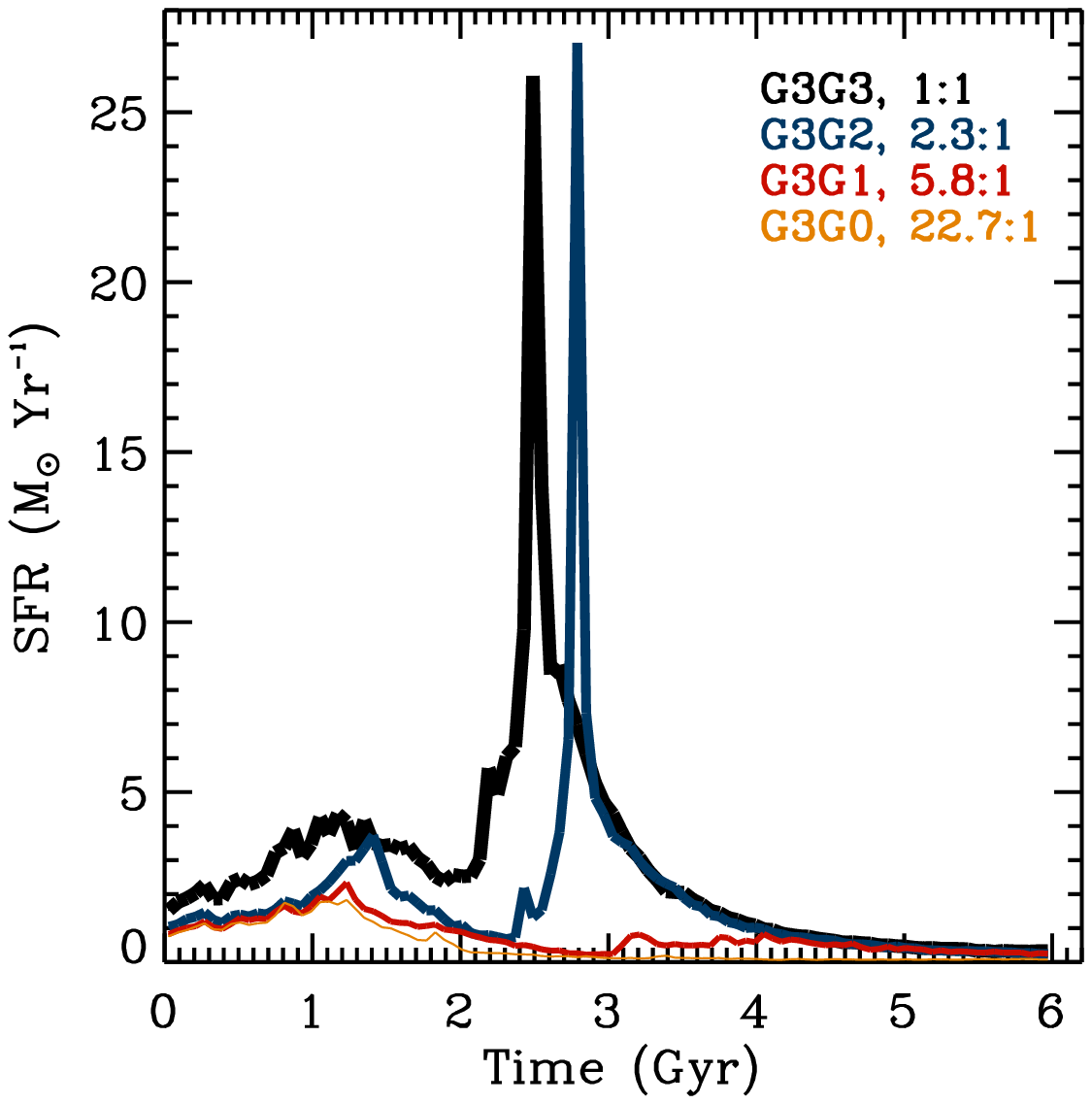}}
\resizebox{8.0cm}{!}{\includegraphics{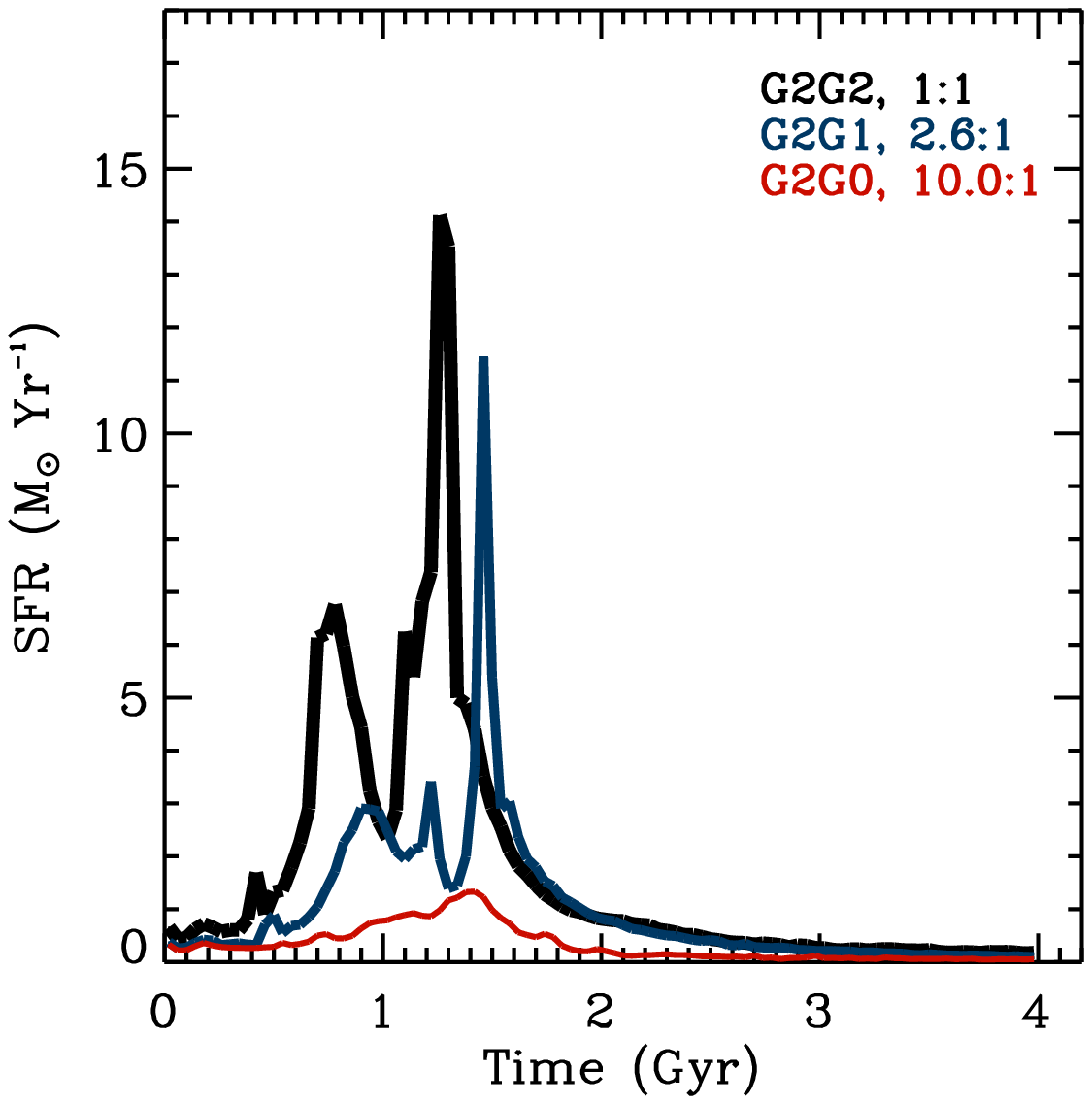}}\\
\resizebox{8.0cm}{!}{\includegraphics{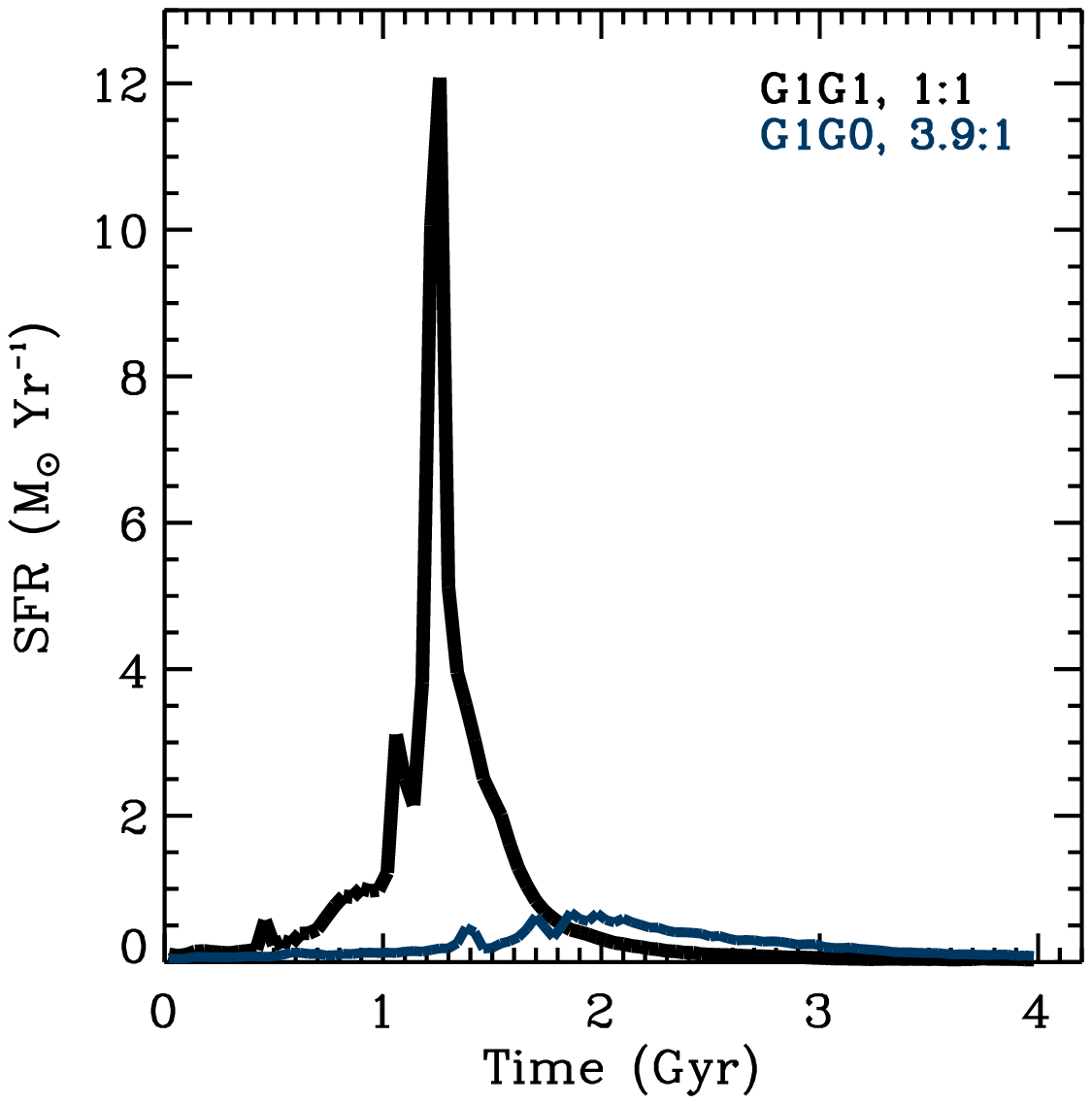}}
\resizebox{8.0cm}{!}{\includegraphics{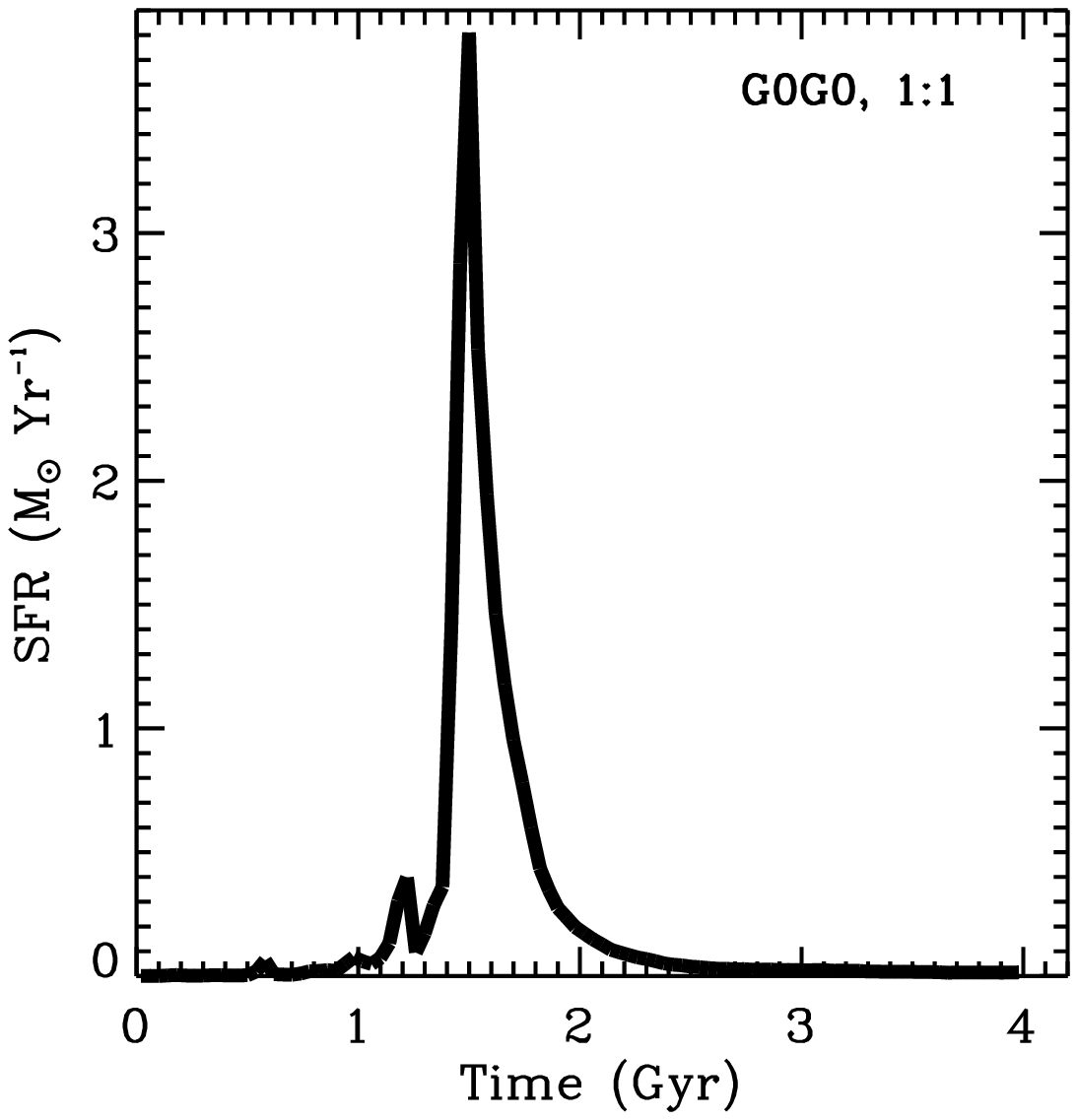}}\\
\caption{Star formation for our fiducial minor merger series.
The total mass ratio is listed next to each run label.  The line
thickness increases with decreasing mass ratio, e.g., the thickest
line in each panel is the 1:1 major merger, while the thinnest line
is the large mass ratio minor merger.  Note
that the vertical axis changes from panel to panel, based upon
the maximum star formation rate for the major merger.  All runs
shown here used the $n0med$ feedback model.
\label{fig:sfr}}
\end{center}
\end{figure*}

The star formation history for all of our fiducial galaxy mergers is displayed in 
Figure~\ref{fig:sfr}.  There are four plots in this figure, one for each of the four primary
galaxy models.  The upper--left panel shows the star--formation rate (SFR) for all
interactions that involve the most massive galaxy G3.  The remaining panels show
all the remaining mergers, grouped by the largest participant galaxy.  Thus each
panels shows the SFR for one major merger and one or more minor mergers, except for
the G0 panel which only has the G0G0 major merger.

The SFRs shown in 
Figure~\ref{fig:sfr} are clearly enhanced compared to the quiescent SFRs shown
in Figure~\ref{fig:sfriso}.  For example, the maximum SFR during the G3G3 
interaction ($\sim26$~\msunyr) is 13 times larger than the summed SFR of two isolated
G3 disks ($\sim1$~\msunyr), the maximum SFR during the G3G2 interaction 
($\sim27$~\msunyr) is 21 times larger than an isolated G3 ($\sim1$~\msunyr) plus an
isolated G2 ($\sim0.25$~\msunyr), and the maximum SFR during the G0G0 interaction
($\sim3.8$~\msunyr) is 3800 times larger than two isolated G0 disks 
($\sim10^{-3}$~\msunyr), however \S\ref{sssec:sffbps} demonstrates that these
factors depend upon the feedback model.

In \S\ref{ssec:psf} we will quantify the relationship between the mass ratio of the
interacting galaxies and the enhanced star formation in more detail, but we note
that this trend fits naturally within a merger--driven scenario for star formation.
In particular, during an interaction gas is stripped of its angular momentum by
bar-like structures during the early stages of the merger or the abundant collisions
and a growing potential well during the messy coalescence.  In both cases the
resulting central concentration of gas fuels a burst of star formation.  Since the
tidal forces associated with the merger generate these effects, their magnitude is expected
to scale with the size of the perturbation.  While this general picture has been
studied in great detail for collisions between equal mass galaxies \citep[e.g.,][]
{MH94minm,MH96,Sp00,Cox06}, we demonstrate here that the star formation clearly
depends on the mass ratio of the interacting galaxies.

\begin{figure}
\begin{center}
\resizebox{8.0cm}{!}{\includegraphics{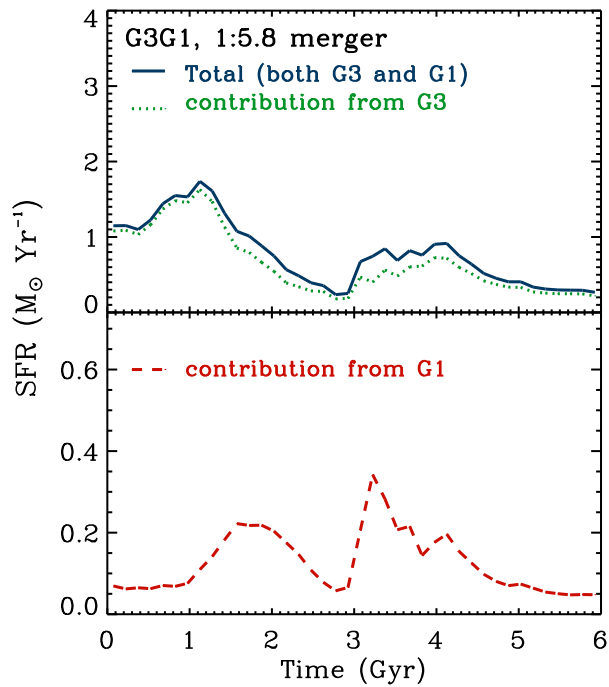}}
\caption{Star--formation rate during G3G1 interaction including the 
contributions from participant galaxies G3 and G1 separately.
\label{fig:sfrg12}}
\end{center}
\end{figure}

We caution that the global SFR, as is shown in Figure~\ref{fig:sfr}, will be dominated
by the largest galaxy when the interacting galaxy mass ratio is large, as in the 
G3G1 and G3G0 mergers.  In this case, the small enhancement in the SFR is only
an indication that the primary, G3, is not tidally disturbed enough to induce
radial inflows of gas and additional star formation.  The less massive satellite, 
however, is expected to experience significant tidal forces and may have a large
enhancement of star formation that is hidden within the global rate.  Such is
the case for G3G1, as shown in Figure~\ref{fig:sfrg12}, where the global SFR is
shown as well as the contributions from both the primary and the satellite.  While
G3, the larger galaxy, constitutes the majority of the overall star formation, its
rate is largely unchanged from the initial stages and eventually decays as a result
of gas depletion.  There is a small enhancement of star formation above the quiescent
G3 from $T\approx3-5$~Gyr coincident with the final coalescence.  In contrast, the
satellite galaxy G1 experiences several bursts of star formation that are $>3$
times its quiescent rate, consistent with merger--driven star formation.

The exact relationship between merger mass ratio and enhanced star formation is
complicated by the fact that the star formation history also depends upon the nature
of the interaction, the structure of the participating galaxies, and the ``sub--grid''
model for star formation and feedback.  The first two dependencies are inherently
physical assumptions and will be investigated further in \S\ref{sec:omods}.  The last
dependency, however, is determined by numerical free parameters that are not well
constrained.  In \S\ref{sssec:sffbps} this issue will be considered further, with
the specific goal of determining which features of the star--formation history provide
a measure of the merger--driven star formation that are insensitive to assumptions
about the feedback model.

\subsubsection{Location of Starburst}
\label{sssec:sfloc}

\begin{figure}
\begin{center}
\resizebox{8.0cm}{!}{\includegraphics{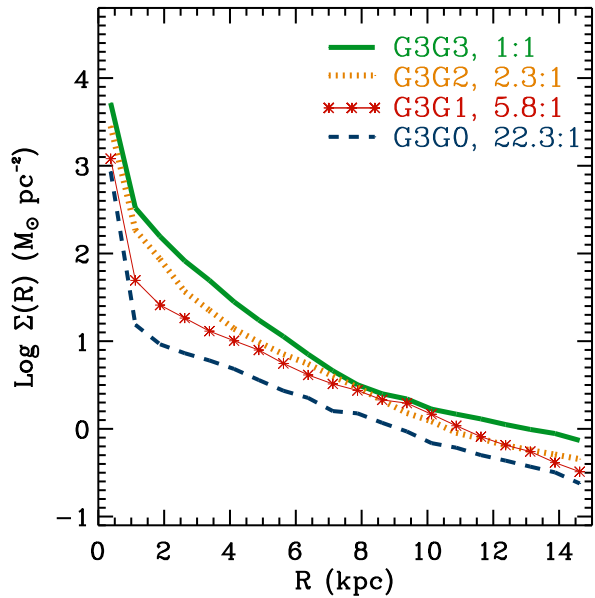}}
\caption{Mass surface density of ``new stars,'' i.e., stars that are
forming during the interaction, for mergers G3G3, G3G2, G3G1, and
G3G0.
\label{fig:prof}}
\end{center}
\end{figure}

One result of merger--driven star formation is a concentration of newly formed 
stars in the galactic center.  Figure~\ref{fig:prof} demonstrates that this is
a generic outcome of our interactions by showing the surface mass density
of ``new stars,'' i.e., stars that form during the interaction, for the
remnants of the G3Gx set of interactions.  Numerous numerical simulations have
previously found similar profiles (though often plotted versus $r^{1/4}$) for
remnants of major mergers \citep[e.g.,][]{MH96,Sp00,Cox06}.  The central
concentrations of new stars are often disjoint from the outer, old--star
profiles, in apparent conflict with the $r^{1/4}$ profile that most ellipticals are
assumed to have.  While this was once considered a problem for the merger
hypothesis, recent high--resolution observations now indicate that such features
exist in nearly all merger remnants \citep{RJ04} and many low--luminosity
elliptical galaxies \citep{Kor07}.

Because the gaseous inflows that produce these central concentrations of new
stars are produced by the gravitational forces arising from the interaction, we
expect the amount of new stellar mass in the galactic center to correlate with
the merger mass ratio.  Indeed, Figure~\ref{fig:prof} demonstrates that the
new star profile within $\sim4$~kpc becomes monotonically steeper with decreasing
merger mass ratio.  The least disruptive mergers, G3G1 and G3G0, have profiles
that are essentially exponential beyond $\sim1$~kpc, while the smaller mass
ratio mergers G3G3 and G3G2 have profiles that are better fit by $r^{1/4}$--type
distributions.

Another way to characterize the distribution of star formation as a result of the
interaction is by simply quantifying the fraction of gas that is driven to within
the central several hundred pc of the primary.  This quantity was an especially
relevant measure of the potential starburst event when numerical simulations
did not include star formation \citep[e.g.,][]{H89,HM95}.  In
these works it was always found that $>35$\% of the gas in the primary was driven
to within the central several hundred pc.  The only exception was when \citet{HM95}
adopted a model in which the interstellar medium was isothermal with a temperature
$>3\times10^5$~K.  We note that we always find that $<35$\% of the gas in the
primary is driven to within 500~pc, even during the G3G3 major merger.  For 
the large mass ratio mergers this percentage is closer to 10\%, which is similar
to the G3 galaxy evolved in isolation.  This trend toward smaller inflows of gas
is another manifestation of the differences between our modeling and what was
performed previously (see more extensive discussion in \S\ref{ssec:compS}).

While it is beyond the scope of the present work, analysis of the
remnant profiles uncovers two interesting avenues for future study.
First, it is intriguing that the new--star profiles of the remnants of
very large mass-ratio mergers are well fitted by a bulge plus
exponential disk profile, suggesting that such minor mergers may be an
efficient mechanism for growing galactic bulges.  Second, at large
($\sim10$~kpc) radii the profiles of the large mass ratio remnants
differ from the G3 evolved in isolation owing to induced star
formation and angular momentum transfer during the early stages of the
interaction \citep[see also][]{Yng07}.

\subsubsection{Dependence on Star Formation and Feedback Parameters}
\label{sssec:sffbps}

\begin{figure*}
\begin{center}
\resizebox{5.3cm}{!}{\includegraphics{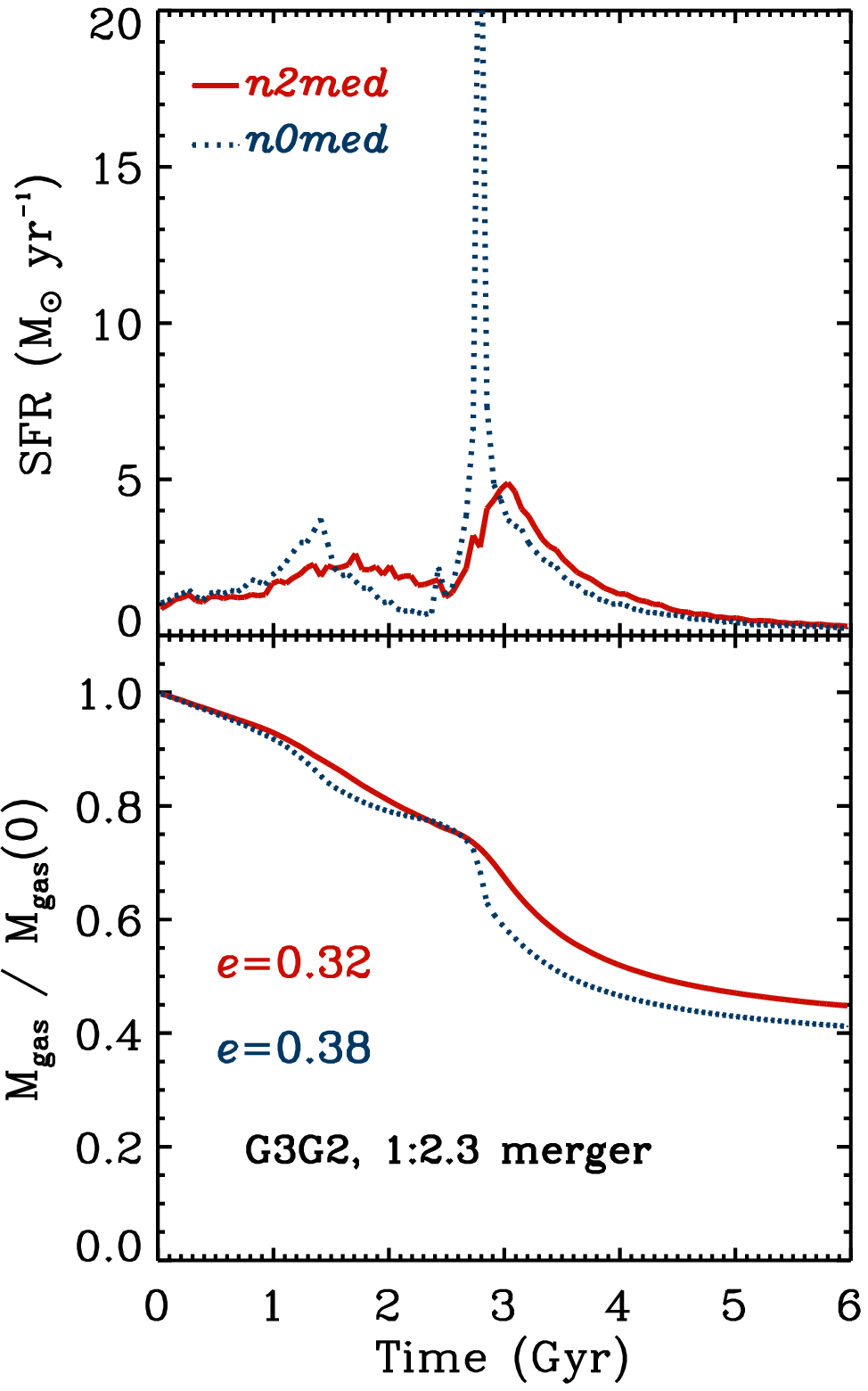}}
\resizebox{5.3cm}{!}{\includegraphics{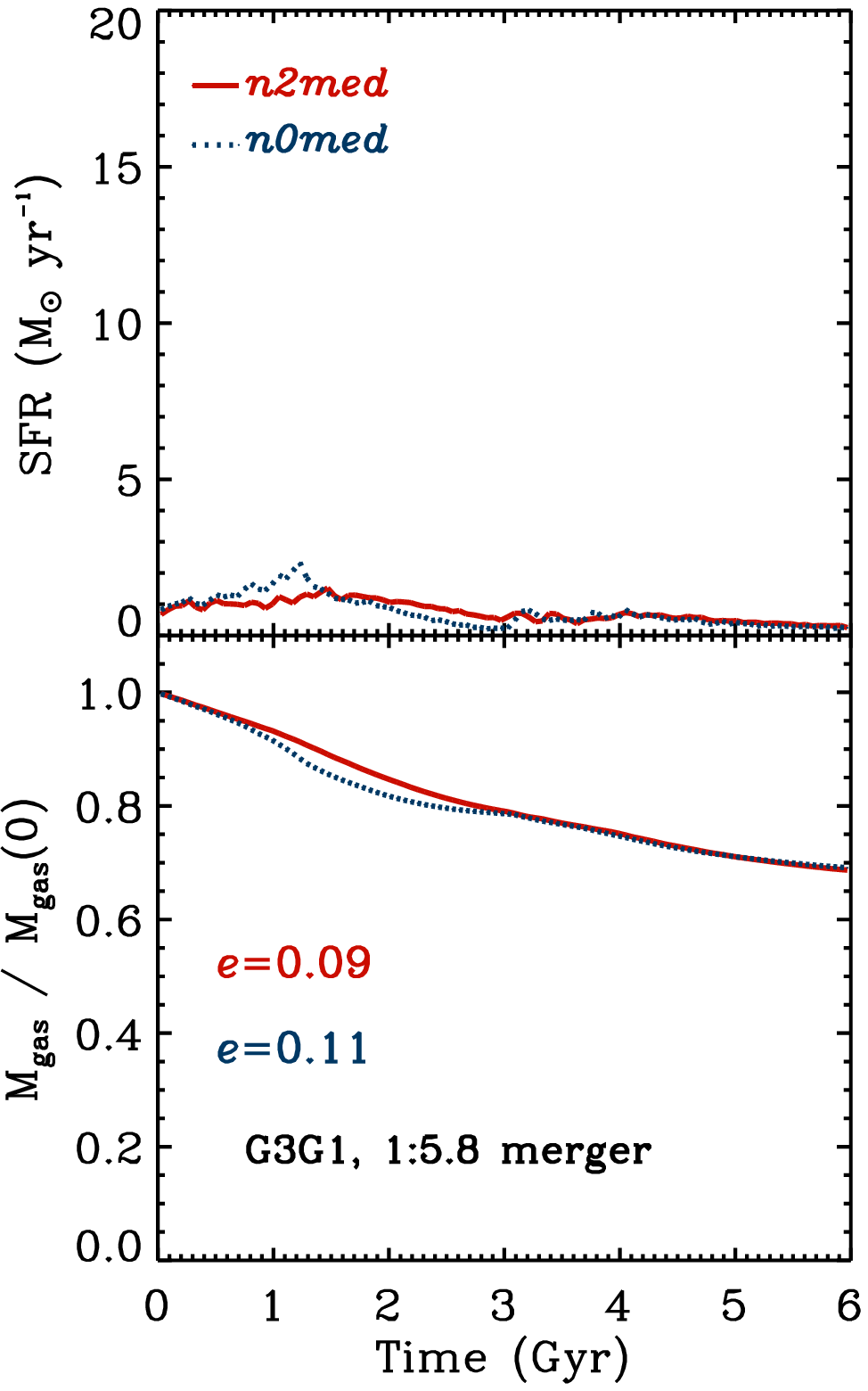}}
\resizebox{5.3cm}{!}{\includegraphics{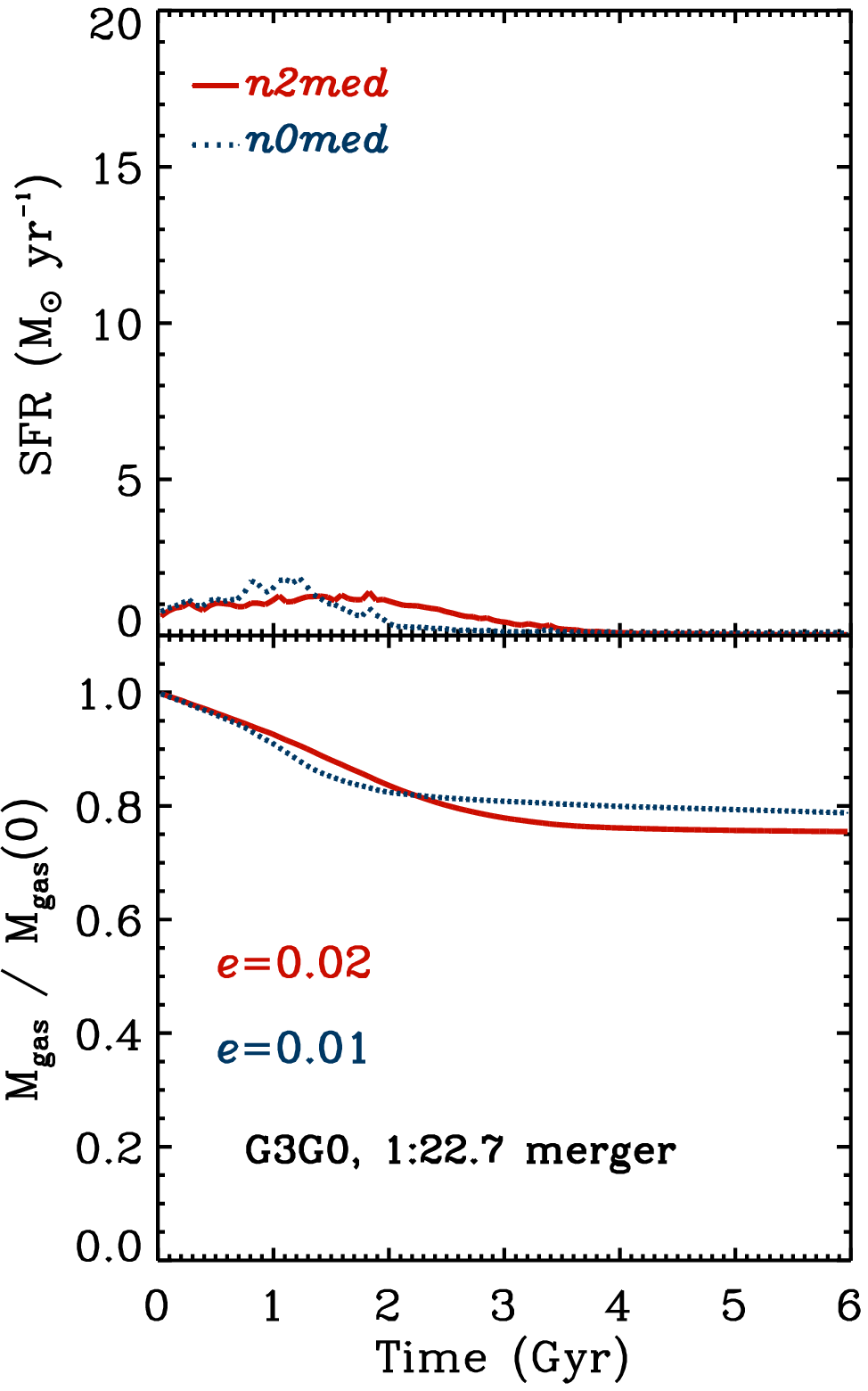}}\\
\caption{Comparison between two parameterization of star
formation and feedback in identical unequal mass mergers.
The top panels show the star-formation rate, and the bottom
panels show the gas consumption.  In general, details of
the star-formation history depend on the feedback model
while the gas consumption remains relatively invariant.
 The burst efficiency (see \S\ref{sssec:be}) of each run
($n2med$ on top, and $n0med$ on the bottom) is provided in
the lower panel.
\label{fig:mincomp}}
\end{center}
\end{figure*}

\begin{figure}
\begin{center}
\resizebox{8.0cm}{!}{\includegraphics{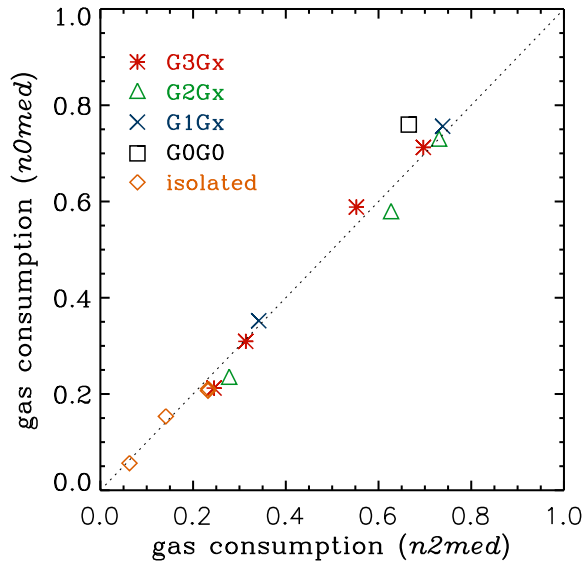}}
\caption{Comparison of the gas consumption during equivalent
simulations performed with two different parameterizations for
feedback, $n0med$ and $n2med$.  The label for the merger
simulations are grouped by the primary.   The isolated
galaxies are identical to the simulations presented in
Figures~\ref{fig:isomorph}~and~\ref{fig:sfriso}.
\label{fig:sffb_comp}} 
\end{center}
\end{figure}

As mentioned previously (\S\ref{sec:sims} and \S\ref{ssec:sfr}), and shown explicitly
by \citet{Cox06} using numerical simulations of major mergers, the 
star-formation history during an interaction depends upon the adopted feedback model.
This is in contrast to the star--formation history of an isolated spiral, which,
as shown in Figure~\ref{fig:sfriso} \citep[and in ][]{Cox06}, is independent of the 
feedback model.
As a result, it is unclear what measure of the star--formation history will provide
a robust characterization of the merger--driven star formation.  To investigate this
further, we resimulated our fiducial set of mergers using the $n2med$
feedback model in addition to the $n0med$ that was presented in  \S\ref{ssec:sfr}.
As shown in \citet{Cox06}, the ``medium''
feedback models are slightly favored because they can maintain a stable gas disk
in an isolated gas--rich Sbc galaxy model and also produce merger--driven starbursts.  The
$n0med$ model treats star--forming gas with an isothermal ($\sim10^{5}$~K) equation
of state while the $n2med$ has a stiff equation of state that restricts the quantity
of gas at very high densities.

Figure~\ref{fig:mincomp} shows the SFR and gas consumption during the G3G2, G3G1,
and G3G0 unequal mass mergers for both the $n0med$ and $n2med$ feedback models.  For all
interactions the peak of star formation begins sooner, has a much larger maximum, and
a shorter duration in the $n0med$ feedback model as opposed to $n2med$.  However, regardless
of the differences in the star-formation history, the gas consumption is very similar,
as shown for the full series of merger simulations in Figure~\ref{fig:sffb_comp}.  This
result motivates us to focus our analysis upon quantities that use the gas consumption
rather than details of the time--dependent star-formation history, since the gas
consumption is invariant to assumptions about the feedback model.  We also note that the
non--equivalence in gas consumption between the two feedback models that is present in
Figure~\ref{fig:sffb_comp} can be used to estimate our errors owing to adopting a
single feedback model.  From the standard deviation in discrepancies we estimate an
error of $\sim0.04$ in the gas consumption owing to the choice of feedback model.
However, we also caution that 3 out of the 4 major mergers have slightly ($<0.1$)
higher gas consumption when adopting the $n0med$ feedback model, and the 2 minor
mergers with the largest mass ratios have higher gas consumption when adopting the
$n2med$ feedback model.  Therefore, it is possible that subtle, yet systematic, trends
exist between alternate feedback models.

As a final comment we note that Figure~\ref{fig:mincomp} demonstrates that the feedback
model can drastically affect the time--dependent star--formation rate, and therefore the
luminosity of the system, even though the total gas consumption is similar.  The varying
luminosity evolution admits the possibility to better constrain the feedback model through
a comparison to the observed distribution of star--formation rates 
\citep[see, e.g.,][]{Noeske07short}.

\subsubsection{The Star--Formation Timescale}
\label{sssec:sfts}

\begin{figure}
\begin{center}
\resizebox{8.0cm}{!}{\includegraphics{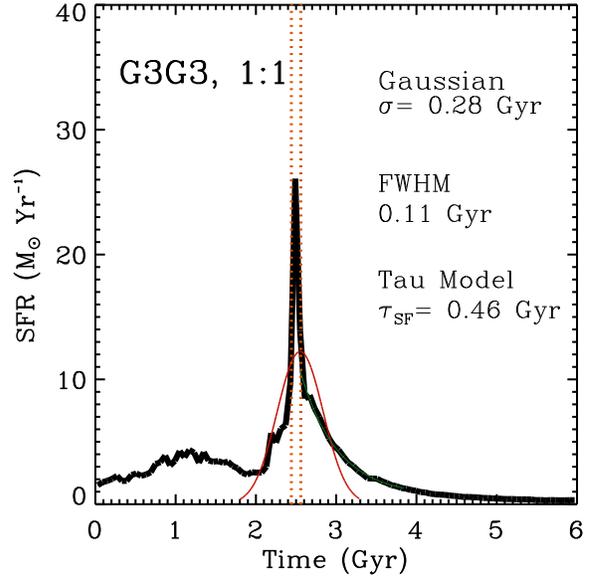}}
\caption{Star--formation rate during the G3G3 major merger
with the $n0med$ feedback parameters (thick solid line).
Overplotted are three characterizations of the merger
timescale, one, a single Gaussian fit to the merger--induced
peak of star formation (thin solid line), two, a decaying
exponential fit to the star--formation rate after the peak
of star formation and, three, the full width at half
maximum (vertical dashed lines).
\label{fig:sfrtime}}
\end{center}
\end{figure}

\begin{figure}
\begin{center}
\resizebox{8.0cm}{!}{\includegraphics{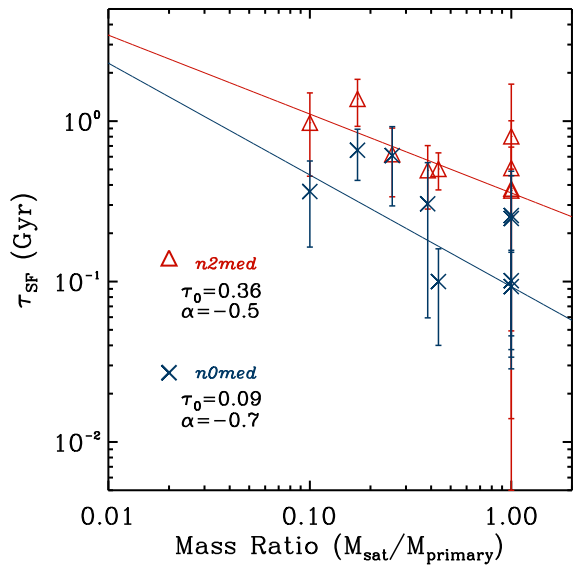}}
\caption{The star formation timescale ($\tau_{\rm SF}$) as a function
of merger mass ratio for our fiducial series of mergers when
simulated with both the $n0med$ and $n2med$ feedback models.
The timescale is the average of the three fits (Gaussian, exponential,
and FWHM) scaled to the exponential decay timescale $\tau$ and the
errors are estimated by the dispersion between the three fit
parameters.  The solid lines are fits to the simulated data using
Equation~\ref{eq:fitsfts}.
\label{fig:sfrts}}
\end{center}
\end{figure}

The previous section demonstrates that each feedback model yields
a unique star formation history, and noted that one difference between
the histories is the star--formation timescale.  Because differences
in the star--formation timescale directly affect the predicted color,
line--strength, and nuclear activity of the interacting pair, future
work may have the ability to discriminate between feedback models based upon
comparisons with observations \citep[e.g.,][]{Bar00,NCA04,WGB06,Gel06,
WG07,Ell07,Bar07}.

As a preliminary step toward this goal, we quantify the starburst
timescale for the mergers simulated here from three different fits to
the starburst events.  The first fit is motivated by the intrinsic
shape of the starburst events as exemplified in
Figures~\ref{fig:sfr}~and~\ref{fig:mincomp}, namely they appear 
Gaussian in shape and their timescale can be characterized by the width
$\sigma$.  The second fit is motivated by many models of galaxy evolution
that describe star formation as a decaying exponential with a timescale
$\tau$, and are therefore often described as ``tau''-models \citep[see,
e.g.,][]{SP99,Hark06,Noeske07short}.
The last method to describe the star formation timescale calculates the
full width at half maximum (FWHM) and therefore avoids any biases
associated with adopting a specified distribution for of the starburst.
Another possibility, which we reserve for future work, is to compile
the full luminosity distribution for the entire merger as has been
done previously by \citet{Patrik} and \citet{Hop06merg}.

As an example of a starburst event, and its corresponding fits, we show
the merger G3G3 in Figure~\ref{fig:sfrtime}.  Printed directly on the
figure are estimates for the starburst timescale that result from each
fit.  The three separate timescale measures can be placed on an equal
footing by noting that the FWHM of a Gaussian distribution is
2.35$\sigma$, and fitting an decaying exponential to one side of a
Gaussian yields $\tau\approx1.1\sigma$.  For this particular case, the
starburst has a distinct peak and extended tails compared to the Gaussian
fit (i.e., a positive Kurtosis).  Therefore, the timescale derived from
the FWHM is much smaller than that derived from the Gaussian fit.  The
starburst is very well-fitted by the tau--model (so much so that the fit
is difficult to see in Figure~\ref{fig:sfrtime}) which yields a
slightly longer estimate of the star formation timescale.

In general, we note that nearly all models that employ the $n0med$ feedback
have peaky star formation, and thus small FWHMs compared to the Gaussian or
exponential estimates.  However, because the star formation events are
typically irregular, we consider the best estimate of the star formation
timescale to be an average between $\tau$, $1.1\sigma$, and $0.47$FWHM, i.e.,
the estimates from the different fits are all scaled to be in terms of the
exponential decay timescale $\tau$.  This also provides a simple estimate of
the error in our fitting procedure from the dispersion among the three
determinations of the timescale.

In Figure~\ref{fig:sfrts} we plot the estimates for the star formation
timescale and the associated errors for the fiducial set of runs as a
function of merger mass ratio for both feedback models.  To characterize
the mass--ratio dependence, we have fit a simple power--law to the mergers
of each feedback model, viz.,
\be
\tau_{\rm SF} = \tau_0 
     \left(\frac{{\rm M}_{\rm sat}}{{\rm M}_{\rm primary}}\right)^{\alpha}.
\label{eq:fitsfts}
\ee
Confirming our visual inspection of the SFRs shown in
Figure~\ref{fig:mincomp}, the $n2med$ model has a much longer star--formation
timescale than the $n0med$ model.  In additional, both models display a
trend for the star--formation timescale to increase with increasing mass
ratio, however the uncertainties are large enough that this trend is not
definitive.  In particular, the goodness of fit is only slightly reduced
if the fit is fixed to have no mass-ratio dependence at all ($\alpha=0$).

\subsection{Parameterizing Star Formation Enhancements During
            Galaxy Interactions}
\label{ssec:psf}

The star-formation histories shown in Figure~\ref{fig:sfr} demonstrate
clear signs of enhanced star formation during the merger.  Furthermore,
\S\ref{sssec:sffbps} argues that this enhanced star formation is
more robustly characterized by the amount of gas consumed, rather than
time--dependent quantities such as the maximum star formation rate, owing
to uncertainties in the adopted feedback model.  However, we must also
consider that the two galaxies which participate in the interaction
would have converted some of their available gas into stars even in
isolation, and therefore the gas consumption itself does not provide a
complete characterization of the merger--driven star formation.

\begin{figure*}
\begin{center}
\resizebox{16.0cm}{!}{\includegraphics{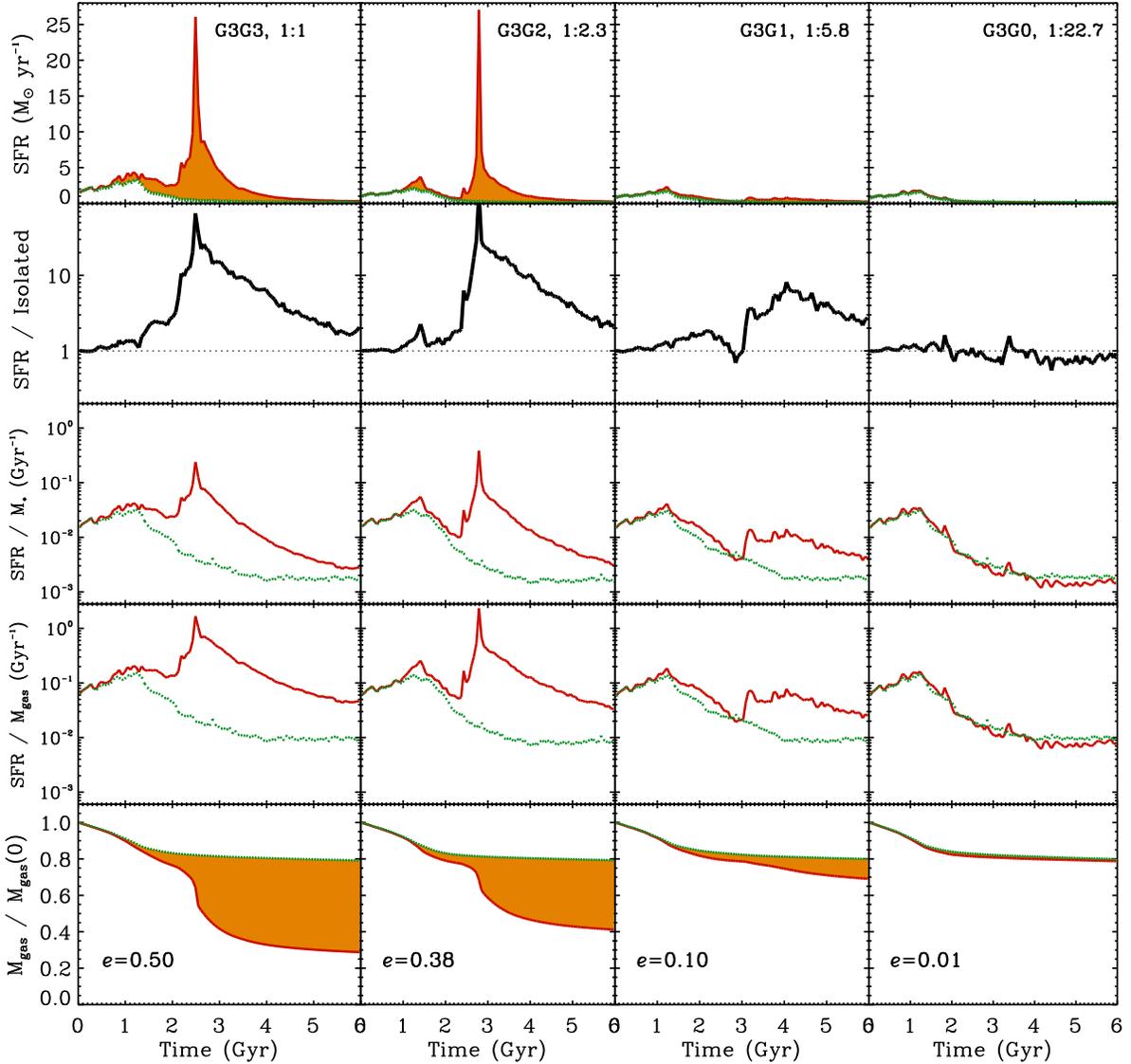}}
\caption{The rows from top to bottom show the star-formation rate
(SFR), SFR normalized by the combined SFR of both isolated galaxies,
the SFR divided by the total stellar mass, the SFR divided by the
total gas mass, and the gas consumption for mergers
G3G3, G3G2, G3G1, and G3G0, from left to right.  Panels show 
quantities during the interaction with a solid (red) line and
for the combined isolated galaxies with a dotted (green) line.
The difference between the two lines is shaded to emphasize the
merger--driven star formation.  All simulations shown employ the
$n0med$ feedback model.  The burst efficiency (see \S\ref{sssec:be})
is printed in the bottom row for each interaction.
\label{fig:sfr2}}
\end{center}
\end{figure*}

A more explicit comparison between the merger--driven star formation and that
of the isolated disks is presented in Figure~\ref{fig:sfr2}, which shows several measures
of star formation for the G3G3, G3G2, G3G1, and G3G0 mergers, in each column,
from left to right, respectively.  The top row presents the star--formation rate
(SFR) for the merger as well as the summed SFR of the primary and satellite galaxies when
evolved in isolation.  For clarity, the difference between the two SFRs is shaded.
The second row shows the ratio of the merger SFR to the combined SFR of the
isolated galaxies, clearly delineating when the rate is enhanced owing to the
merger.  The third row plots both SFRs divided by the stellar mass, which is also the
inverse of the stellar--mass doubling time, and is sometimes termed the ``specific SFR.''
The fourth row plots the SFRs divided by the gas mass, which is the inverse of the 
gas consumption timescale.  Finally, the bottom row presents the gas consumption of
both the merger and the combined isolated galaxies with the difference shaded.

There are several features present in Figure~\ref{fig:sfr2} that deserve more
discussion.  First, reinforcing the notion first presented in Figure~\ref{fig:sfr},
there is a clear correlation between interacting galaxy mass ratio and merger--driven
star formation.  Significant episodes of star formation only occur for mergers between
galaxies of nearly equal mass.  For mergers between galaxies with a large mass ratio,
such as G3G0, the cumulative star--formation history is nearly indistinguishable from
the primary G3 evolved in isolation.

By comparing the star--formation history of G3G3 to that of G3G2, we uncover another
interesting feature of Figure~\ref{fig:sfr2}.  Namely, these two interactions have
similar peak levels of star formation, even though the total amount of gas consumed is
much less for the higher mass ratio G3G2 interaction.  This feature of the
star--formation history emphasizes that caution needs to be exercised when attempting
to quantify the merger--driven star formation.

As a final comment, we note that all models except G3G0 show elevated star formation
at the end of the interaction ($T\approx6$~Gyr) compared to the primary G3 evolved in
isolation, as indicated by the normalized SFR shown in the second row of 
Figure~\ref{fig:sfr2}.  Therefore, most measures of the the merger--driven
star formation are subject to uncertainties that depend upon the duration over
which the simulations follow the merger (if the simulation was run for long
enough, all galaxies would consume essentially all of their initial allotment of gas).  In practice,
these errors are quite small, which we estimate to be $\sim0.03$ from the differential gas 
consumption (bottom row of Figure~\ref{fig:sfr2}) if the simulation is followed for
1~Gyr prior to, or later than, the current 6~Gyr we adopt as a standard.

\subsubsection{Burst Efficiency}
\label{sssec:be}

\begin{figure}
\begin{center}
\resizebox{8.0cm}{!}{\includegraphics{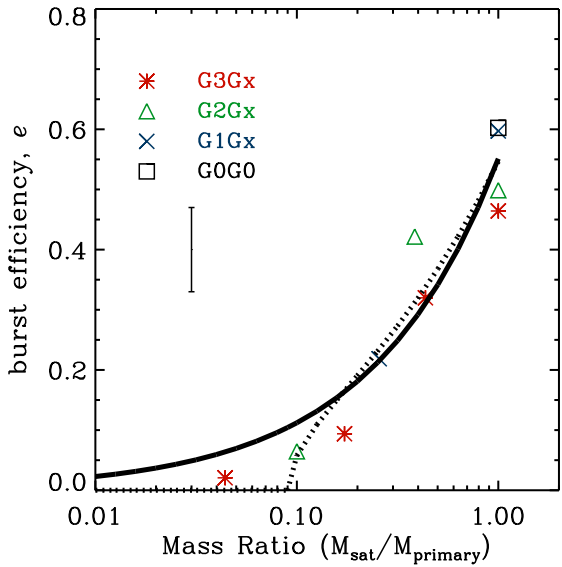}}\\
\caption{Burst efficiency $e$ (defined in Eq.~\ref{eq:definebe}) for
all fiducial merger simulations.  Interactions in which G3 was the largest
progenitor, i.e., G3Gx, are shown with a (red) asterisk, while G2Gx
mergers are (green) open triangles, G1Gx are (blue) X's, and G0G0 is
an open (black) square.  The solid line is the best--fit relation using
Eq.~\ref{eq:be}, and the dotted line is the best--fit relation using
Eq.~\ref{eq:bez}.  The vertical line below the key shows our estimated
error in burst efficiency of $\pm$0.07 owing to uncertainties
in the feedback model and the duration in which the simulation is
followed.
\label{fig:bestd}}
\end{center}
\end{figure}

In order to quantify the merger--driven star formation, we wish to define a
simple, useful, and robust parameterization of the star formation owing
solely to the interaction.  To this end, we are motivated by the discussion
of \S\ref{sssec:sffbps} to introduce the ``burst efficiency''
$e$, as
\be
\begin{array}{lcl}
e & = &  ({\rm fraction~of~gas~consumed~during~interaction})~- \\
  &   &  ({\rm fraction~of~gas~consumed~by~constituent} \\
  &   &   {\rm galaxies~evolved~in~isolation,~during~the} \\
  &   &   {\rm same~time~period}).
\end{array}
\label{eq:definebe}
\ee
The burst efficiency is the fractional amount of stars that are formed (or
equivalently, the gas that is consumed) during the merger that would not
have formed in isolation.
Figure~\ref{fig:bestd} shows the burst efficiency for our fiducial merger
series and reaffirms what was implicit in Figures~\ref{fig:sfr} and 
\ref{fig:sfr2}; namely, merger--driven star formation is only significant
during mergers where the participating galaxies are comparable in mass.

The burst efficiency appears to be a smoothly decreasing function of the
merger mass ratio and its parameterization may be useful for future studies
of galaxy formation.  In fact, such a parameterization has already been
introduced by \citet*[][see their Fig.~1]{SPF} based upon a much smaller set
of data \citep[the simulations of][]{MH94minm,MH96}, and used in their
semi-analytic models.  Following \citet{SPF}, we adopt the following form
for the burst efficiency:
\be e = e_{1:1} 
     \left(\frac{{\rm M}_{\rm sat}}{{\rm M}_{\rm primary}}\right)^{\gamma}
\label{eq:be}
\ee
where $e_{1:1}$ is the burst efficiency for equal mass mergers, M$_{\rm sat}$
and M$_{\rm primary}$ are the mass of the satellite and primary, respectively,
and $\gamma$ fixes the mass ratio dependence.  Performing a least-squares fit
to our entire set of fiducial mergers we find $e_{1:1}$=0.55 and $\gamma$=0.69.
This fit is shown as a solid line in Figure~\ref{fig:bestd}.

Owing to the relatively small star-formation enhancement in large mass ratio
mergers (M$_{\rm sat}$/M$_{\rm primary}<0.2$), and the systematic overestimate
in the regime provided by the best--fit Eq.~\ref{eq:be}, we also consider an alternate form
for the burst efficiency that is fixed to zero when the mass ratio
M$_{\rm sat}$/M$_{\rm primary}$ is below $e_0$ and is
\be  e = \left(\frac{{\rm M}_{\rm sat}}{{\rm M}_{\rm primary}}
         - e_0\right)^{\gamma}, 
\label{eq:bez}
\ee
when the mass ratio M$_{\rm sat}$/M$_{\rm primary}$ is greater than $e_0$.
Performing a least squares fit to our fiducial mergers yields $e_{1:1}$=0.56,
$\gamma$=0.50, and $e_0$=0.09, although the reduced $\chi^2$ is nearly unchanged
if $e_0$ is manually set to anything less than 0.11 (including 0, in which case the
fit is identical to Eq.~\ref{eq:be}).  In other words, our fiducial set of
mergers are consistent with there being no enhancement of star formation below a
mass ratio of 9:1, although the data doesn't necessarily require this.  The best fit
to Eq.~\ref{eq:bez} is shown in Figure~\ref{fig:bestd} as a dotted line.

A number of other fitting formula are also possible for the burst
efficiency, such as single and broken power-laws, and various polynomials,
however with the small number of data points and the
associated uncertainties in calculating the burst efficiency, no one
formula was statistically better than any other formula.  Hence, we
adopt Eq.~\ref{eq:be} to describe the burst efficiency.

For comparison, the \citet{SPF} fit to the Mihos \& Hernquist data using
Eq.~\ref{eq:be} yielded $e_{1:1}=0.75$ and $\gamma=0.18$, when the progenitor
disks did not include a stellar bulge, and $e_{1:1}=0.75$ and $\gamma=1.16$ 
when they did.   The consistency of $e_{1:1}$ indicates that the presence of a
bulge does not affect the burst efficiency of major mergers.  We also note
that a value of 0.75 for $e_{1:1}$ is about 50\% larger than
our preferred value, indicating that the models employed here are less efficient at
turning gas into stars than those of Mihos \& Hernquist.  In fact, this difference
is similar to that found previously by \citet{Cox06} who showed that the feedback
model and the newer version of SPH employed produce these changes.  Our value
of $\gamma$ is between the \citet{SPF} value, a  trend that is likely driven
by the fact that our primary galaxy G3 has a bulge--to--disk ratio that is
lower than in their model.  We will attempt to explore
these dependencies in more detail in the following section.

\section{Additional Merger Simulations}
\label{sec:omods}

The previous section detailed the outcome of 10 mergers, including 6 
unequal mass mergers, between our four galaxy models.  In this section
we perform a number of additional merger simulations in order to 
explore several interesting, and particularly relevant regions of parameter
space; the merging orbit and the structure of the progenitor disk,
specifically the extent of the gaseous disk, the bulge-to-disk ratio,
and the gas fraction of the progenitor disk.

\subsection{Variations in Orbit}
\label{ssec:morb}

\begin{figure*}
\begin{center}
\resizebox{16.0cm}{!}{\includegraphics{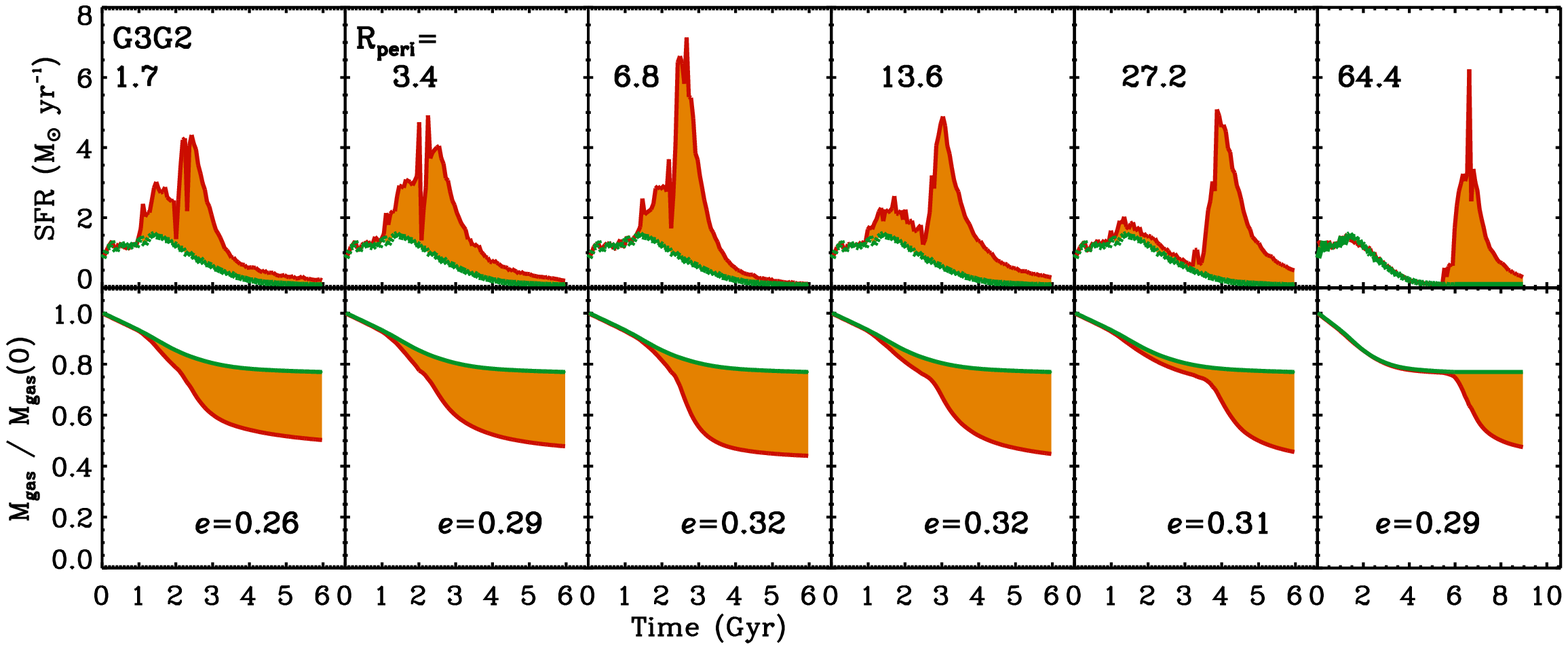}}
\resizebox{16.0cm}{!}{\includegraphics{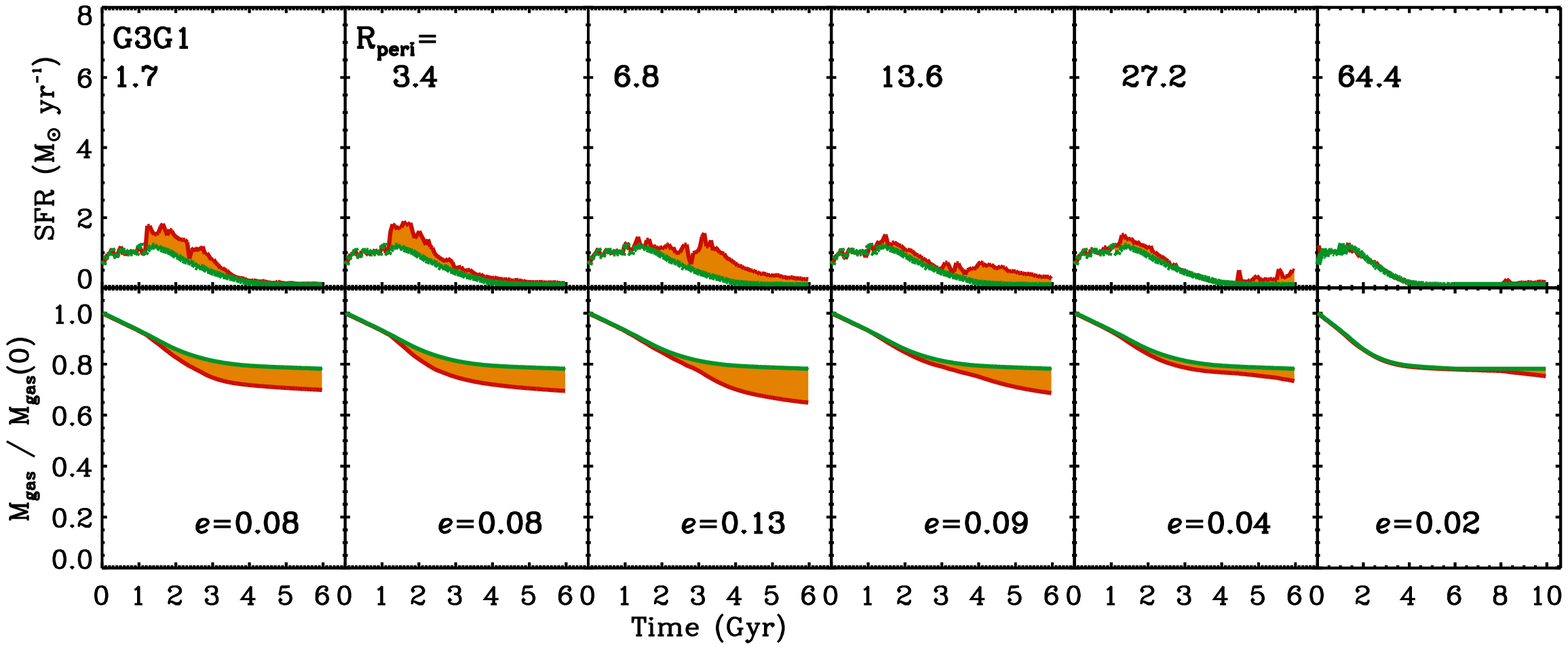}}
\caption{Star--formation history and gas consumption for the G3G2
(top) and G3G1 (bottom) unequal mass interactions when merged on
orbits that have varying amounts of angular momentum, as set by
the pericentric distance R$_{\rm peri}$, which is listed in the
top of each column in units of kpc.  All merger orbits are inclined
by $30^\circ$ and nearly parabolic (eccentricity 0.95).  The fiducial
encounters all assumed R$_{\rm peri} = 13.6$~kpc when the primary was
G3.  The burst efficiency is listed in the bottom of each column.  All
mergers use the $n2med$ feedback model.
\label{fig:sfrinfo_orb}}
\end{center}
\end{figure*}

\begin{figure*}
\begin{center}
\resizebox{16.0cm}{!}{\includegraphics{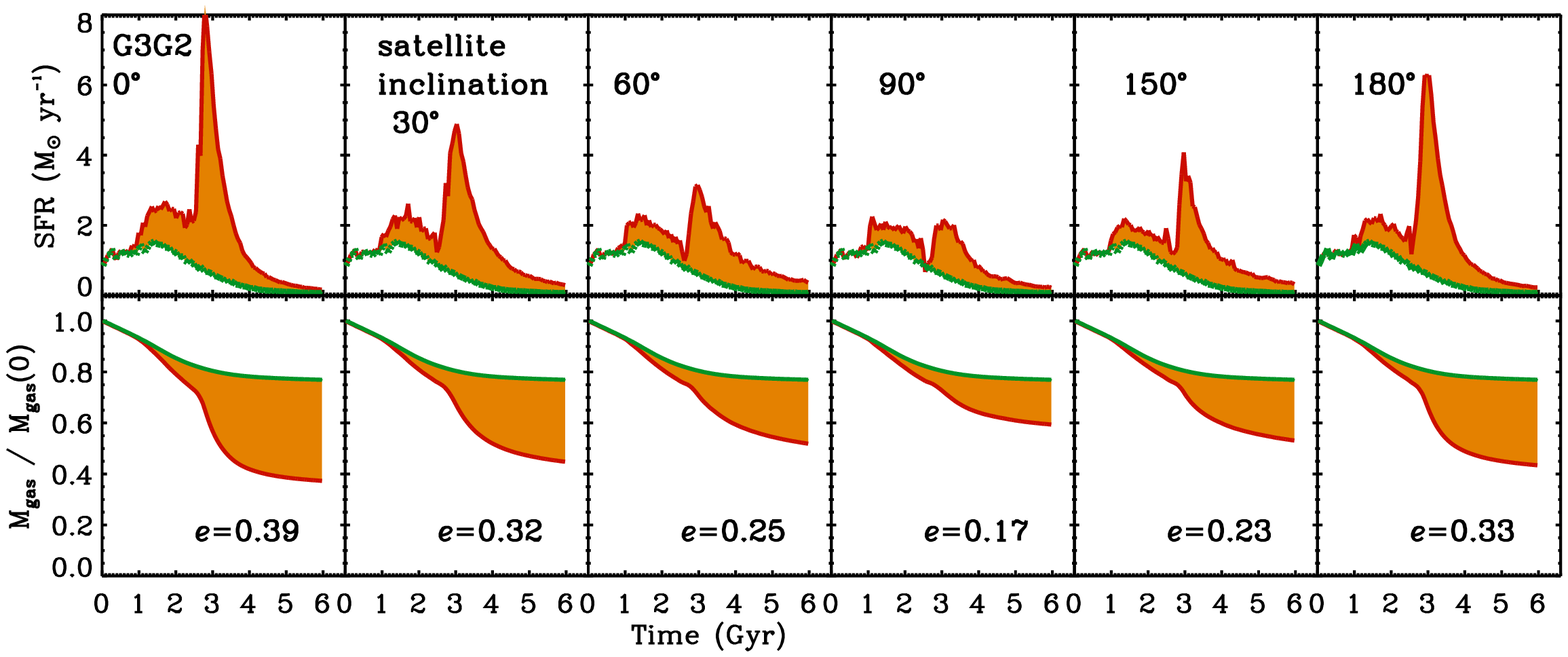}}
\resizebox{16.0cm}{!}{\includegraphics{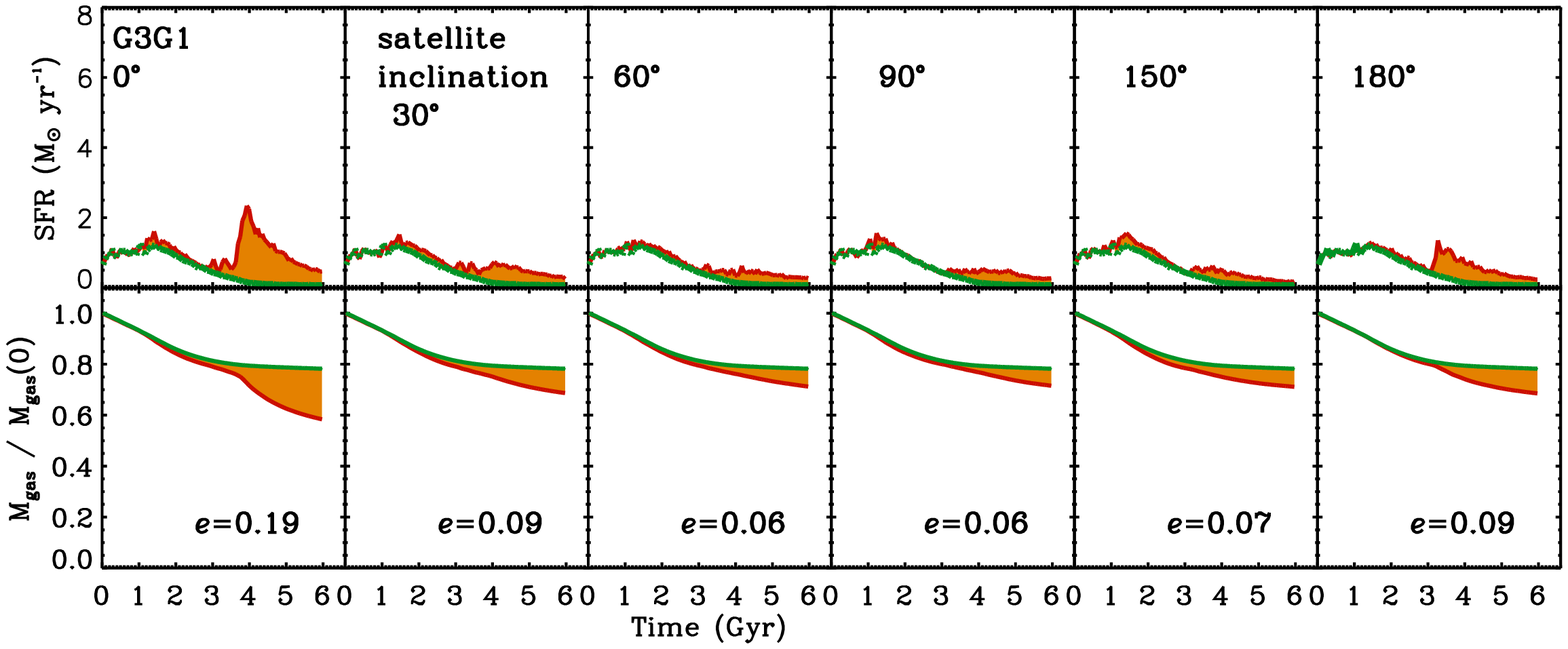}}
\caption{Star--formation history and gas consumption for the G3G2
(top) and G3G1 (bottom) unequal mass interactions when merged on
orbits that have varying satellite orientations, from $0^\circ$ (a
perfectly prograde encounter) to $180^\circ$ (a perfectly retrograde
encounter).  The other parameters of the merging orbits are similar
to the fiducial case, i.e., they are nearly parabolic (eccentricity
0.95) and have moderate angular momentum (R$_{\rm peri} = 13.6$~kpc).
Note that the fiducial set of mergers were inclined by 30$^\circ$.
The burst efficiency is listed in the bottom of each column.
All mergers use the $n2med$ feedback model.
\label{fig:sfrinfo_ori}}
\end{center}
\end{figure*}

As outlined previously (\S\ref{sec:intro} and \S\ref{ssec:sfr}), the enhancement of
star formation during galaxy interactions originates from tidal forces that attend
the merger.  In particular, the close passage between the interacting galaxies
produces bar--like structures in the {\it stellar} disk that torque gas into
the galaxy center.   Figure~\ref{fig:sfrg12} shows that this occurs in both the
primary and the satellite.   The nearly prograde mergers that we have tracked up to
this point produce a relatively strong bar owing to resonances between the
orbits of stars in the primary disk and the passage of the satellite galaxy.
However, alternative orbits and orientations of the interaction will affect
the strength of the tidal force as well as the resonances excited within
the disk, and therefore the resulting starburst.

To determine how the star-formation history depends upon the orbital angular
momentum and disk orientation of the merging galaxies we have run two series
of interactions using the G3G2 and G3G1 mergers.  In the first series, we 
systematically change the orbital angular momentum which, in practice, is
modulated by the pericentric distance R$_{\rm peri}$.  All other parameters
are identical to the fiducial mergers.  The star-formation histories and gas
consumption fractions resulting from this set of mergers are shown in
Figure~\ref{fig:sfrinfo_orb}.  For both the G3G2 and G3G1 interactions the
maximum gas consumption occurs when R$_{\rm peri}=6.8$~kpc, or roughly 
($6.8/2.85\approx$) 2.4 times the stellar scale radius of G3.  The increasing gas
consumption for more direct interactions supports the connection between tidal
forces and merger-driven star formation.  We also note that the two interactions
with R$_{\rm peri}$ less than 6.8~kpc consume slightly less gas than orbits with
larger R$_{\rm peri}$.  This trend is likely a result of the disruptive nature of
nearly head--on collisions as well the difficulty of capturing shock--induced
star formation within SPH (although recent star formation models have been
formulated to address this shortcoming; \citet[see, e.g.,][]{B04}).

The second series of runs systematically alters the satellite orientation.  While
the fiducial set of mergers assumes that the orbital plane was inclined by 30$^\circ$
with respect to the disk of the primary galaxy, this series varied the
inclination.  A range of mergers are performed in which the inclination is prograde
($0^\circ$), polar ($90^\circ$), and retrograde ($180^\circ$), as well as several
in between.  Figure~\ref{fig:sfrinfo_ori} shows the resulting star formation and 
gas consumption for these runs.  As expected, mergers in which the orbit of the 
secondary is aligned with the primary disk, i.e., prograde orientations, produce
the strongest tidal responses in the disk and therefore the largest bursts of star
formation.

It is likely that similar results would follow from other orbital explorations, such
as the initial separation between the two galaxies, as well as the energy of the
initial orbit.  Interestingly, the largest starburst event in the series of major
merger interaction studied by \citet{Thesis} was a zero net angular momentum orbit.  In
short, we expect any factor that efficiently strips gas of its angular momentum, 
including collisions, or increasing the tidal coupling between the satellite orbital
motion and stars in the disk of the primary will lead to the most substantial
bursts of star formation.

\subsection{Variations in the Extent of Gaseous Disk}
\label{ssec:gasd}

\begin{figure}
\begin{center}
\resizebox{8.0cm}{!}{\includegraphics{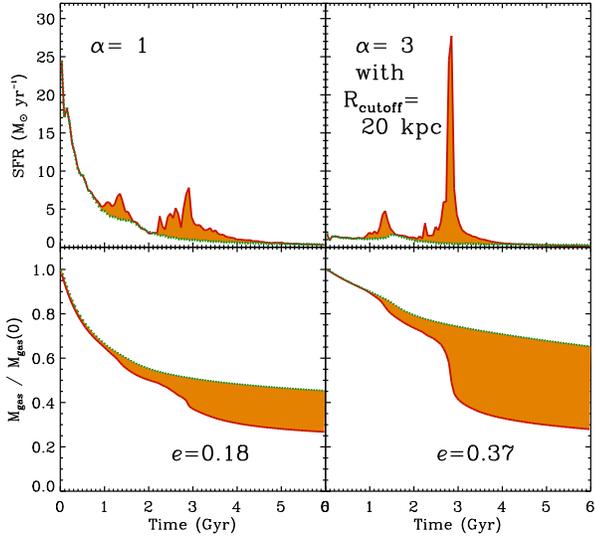}}\\ 
\caption{Star formation and gas consumption for two G3G2 mergers.  The
case on the left is similar to the fiducial interaction except the 
primary G3 galaxy model now has $\alpha=1$, where $\alpha$ is the
multiplicative factor that sets the scale radius of the gaseous disk
with respect to the stellar disk (see \S\ref{sec:ics}).  The case on
the right is also similar to the fiducial interaction except the primary
G3 galaxy model has $\alpha=3$ and all gas beyond 20~kpc has been
removed from the initial disk.  The burst efficiency is indicated in the
bottom panel for each case.  Both mergers are run using the $n0med$
feedback model.
\label{fig:sfr_eg}}
\end{center}
\end{figure}

Motivated by observations \citep[see, e.g.,][]{BvW94}, the galaxy models
that form the basis for this work assume that the distribution of gas in
the disk is more extended than the stars.  Specifically, the gas has an
exponential scale radius $\alpha=3$ times that of the stellar disk.  As
a result, there is a significant quantity of gas at large radii.  This 
extended gas does not significantly contribute to the nuclear starburst.
Instead, the merger remnant contains a large quantity of gas in
both a shock--heated hot phase \citep{Stck97,Cox04,Cox06x} as well as in
a cold star-forming disk. 

In order to discern how the assumed distribution of gas influences the 
burst efficiency, we have run a number of tests that alter the distribution
of cold gas in the primary G3 galaxy.  Figure~\ref{fig:sfr_eg} shows an
example of our typical result, namely, the extended gas distribution leads
to an increased estimate of the burst efficiency.  For the case shown in
Figure~\ref{fig:sfr_eg}, the fiducial G3 galaxy model has been altered to
have $\alpha=1$, while keeping the remaining parameters, including the total
gas mass, unchanged.  In short, the vigorous star formation in the isolated
primary consumes a larger fraction of the available gas and results in a
lower burst efficiency.

In additional experiments we merged models with the fiducial G3 distribution
of gas ($\alpha=3$), however all gas beyond a cutoff radius (R$_{\rm cutoff}$)
is removed.  In essence, these galaxy models have much less gas, yet the inner,
star--forming gas distribution is unchanged.  As shown in Figure~\ref{fig:sfr_eg},
the resulting star--formation histories are unchanged.  Even though the
fractional gas consumption increases markedly (compared to the second column in
Figure~\ref{fig:sfr2}), owing to a corresponding increase in the gas consumption
of the isolated system, the resulting burst efficiencies are nearly identical.
These experiments indicate that the burst efficiency is predominantly
a function of the density distribution of gas in the progenitor systems
rather than purely the spatial distribution of gas.

\subsection{Variations in Bulge-to-Disk Ratio}
\label{ssec:bdrat}

One of the most significant results from the work of
\citet{MH94minm,MH96} was the discovery that the internal structure of
the primary galaxy can strongly influence the efficiency of the
starburst.  Specifically, these works showed that a massive stellar
bulge stabilizes the disk against tidal perturbations and suppress
strong inflows of gas that lead to starbursts.  In the case of minor
mergers, the presence of a bulge may eliminate the merger--driven
starburst completely, while during a major merger the bulge may simply
delay the starburst until the final coalescence of the two galaxies.

The absence of any discernible burst of star--formation for the G3G1 and G3G0
minor mergers presented in \S\ref{sec:mergers} support the notion that the
presence of a stellar bulge can suppress merger--driven star formation
when the mass ratio is large.  Furthermore, this burst suppression persists
even though the bulge adopted for the fiducial G3 model is only 20\% the
mass of the stellar disk, i.e., the bulge--to--disk ratio (B/D) is 0.2, or
about 50\% smaller than in the models of Mihos \& Hernquist where B/D= 0.33.
In order to provide a more direct comparison to their work and further 
assess the effects of bulge mass on the merger--driven starburst, we
have run a number of additional interactions where the bulge mass of the
fiducial G3 model is
altered such that there is no bulge at all (B/D= 0.0), or where the
bulge mass is increased by a factor of 2.5 (B/D= 0.5).  All other parameters,
including the fiducial merging orbits and the satellite galaxies (G2, G1, 
and G0), are left unchanged.

\begin{figure}
\begin{center}
\resizebox{8.0cm}{!}{\includegraphics{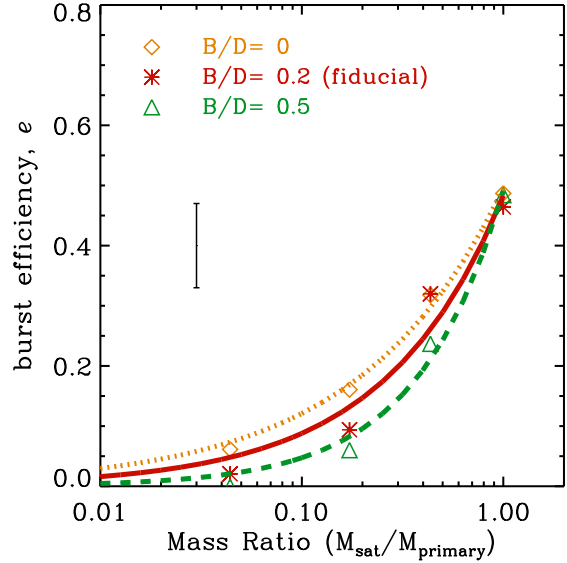}}\\
\caption{Similar to Figure~\ref{fig:bestd} except here the burst
efficiency is shown for three series of runs with different values
of the bulge--to--disk ratio (B/D) in the primary G3 galaxy.
\label{fig:be_bd}}  
\end{center}
\end{figure}

Figure~\ref{fig:be_bd} shows the burst efficiency as a function of merger mass
ratio for our series of mergers in which the bulge mass of the G3 primary is
varied.  As expected, there is a systematic correlation for large mass
ratio mergers to have smaller burst efficiencies when the bulge mass increases.
For example, the bulgeless G3G1 (5.8:1) merger 
has a burst efficiency of 0.16, which is three times larger than the burst
efficiency for the model with the most massive bulge (B/D= 0.5).  We also note
that the burst efficiency is insensitive to the bulge mass when the merger mass
ratio is near unity, in agreement with \citet{MH96}.

Using Eq.~\ref{eq:be}, the best fit to each series of B/D mergers is determined
and overplotted in Figure~\ref{fig:be_bd}.  All fits have the identical value
of $e_{1:1}$, 0.49, a result of the constant burst efficiency during major mergers.
The value of $\gamma$, which sets the mass ratio dependence, is 0.61, 0.74, and
1.02 for the series with B/D= 0.0, 0.2 (the fiducial), and 0.5, respectively.
It should be pointed out that the best fit for the fiducial series is slightly
different than that reported in \S\ref{sssec:be} because here only the G3Gx series
is analyzed.

Even though the qualitative relationship between bulge mass and burst efficiency
is in agreement with \citet{MH94minm}, their bulgeless minor merger produced a
significant burst of star formation.  The result is a burst efficiency ($\sim0.7$)
that is much larger than our bulgeless series.   Actually, this value is larger than
{\it any} of our interactions, major or minor.  This discrepancy was also noted in
\S\ref{sssec:be}, and is likely a result of three differences between their 
modeling and ours.

First, the feedback model and the newer entropy--conserving version of SPH
employed here both result in less intense episodes of merger--driven star
formation \citep{Cox06}.  Second, the models employed by Mihos \& Hernquist 
adopt a circular orbit for the satellite, which increases the tidal coupling 
to stars in the disk and produces a larger response.  Lastly, the large burst
efficiency found by Mihos \& Hernquist is inflated by the very inefficient 
star formation assumed to occur in their quiescent disk.  Such levels of star
formation appear to be insufficient to match more recent observations
\citep{Kenn98}.

\begin{figure}
\begin{center}
\resizebox{8.0cm}{!}{\includegraphics{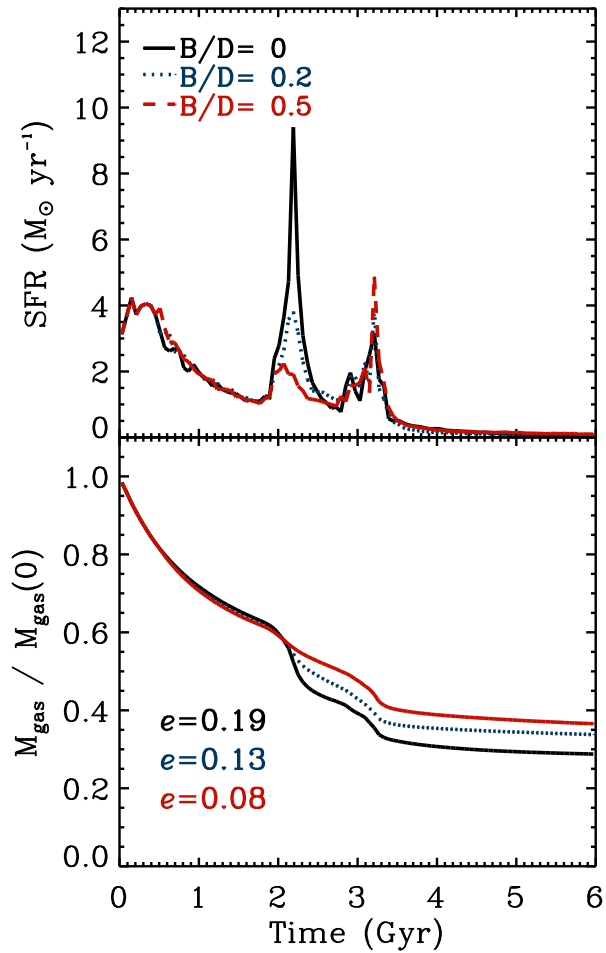}}\\
\caption{Star formation during unequal mass interactions as a function of 
primary bulge mass.  The interaction is similar to the fiducial G3G1, except
the orbit is co--planar ($0^\circ$) and has less angular momentum (see text)
in order to produce the maximum starburst.  All runs are performed with the
the $n0med$ feedback model.
\label{fig:sfr_bd}}
\end{center}
\end{figure}

While these three arguments outline why our burst efficiencies are more modest ---
and we believe, more accurate --- than prior calculations, the previous sections of this
paper suggest that a number of the parameters for our fiducial encounters are
sub--optimal at producing the largest merger--driven star formation event (and
hence burst efficiency).  We therefore performed a small number of additional
simulations using a bulgeless version of our fiducial G3G1 merger, only we placed
the satellite G1 on a co--planar ($0^\circ$), close passage (R$_{\rm peri}= 6.8$~kpc)
orbit in order to maximize the merger--induced starburst (this orbit was motivated by
the results of \S\ref{ssec:morb}).

The star--formation history during the interaction with the maximum burst
efficiency, as determined from our small G3G1 parameter search, is shown in 
Figure~\ref{fig:sfr_bd}.  We have also shown corresponding interactions when the
modified bulgeless G3 model has bulge--to--disk ratios of 0.2, similar to the
fiducial G3 model, and 0.5.  Even with the large degree of parameter manipulation
the maximum burst efficiency for the bulgeless run is only 0.19.  While this is
$\sim20$\% larger than the the bulgeless G3G1 in the fiducial encounter, and about
$\sim4$ times the fiducial G3G1 merger, it is still far below the $\sim0.70$ found by
\citet{MH94minm}.

\subsection{Variations in Gas Fraction}
\label{ssec:gasf}

\begin{figure}
\begin{center}
\resizebox{8.0cm}{!}{\includegraphics{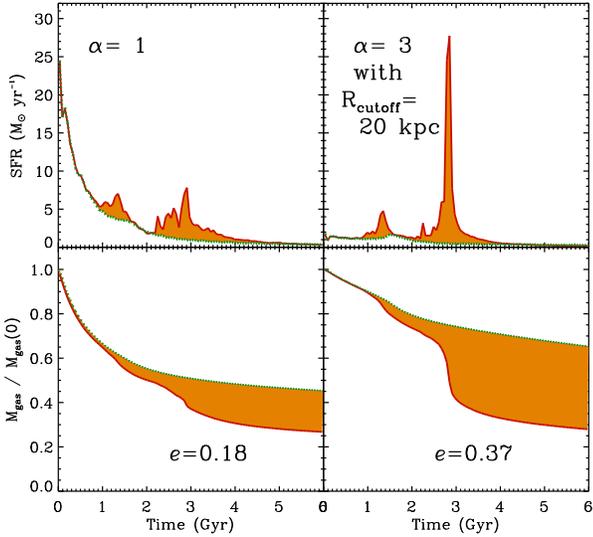}}\\
\caption{Star formation history and gas consumption for the
fiducial G3G2 merger and two similar interactions where the primary
G3 galaxy has a larger gas fraction.  As with Figure~\ref{fig:sfr2},
the difference between the merger and isolated evolution is shaded
for clarity.  The burst efficiency is displayed in the lower--right
of the bottom panel.  We note that the simulation here were run with
the $n2med$ feedback model while the mergers shown in
Figure~\ref{fig:sfr2} employed the $n0med$ model.
\label{fig:sfr_f}}
\end{center}
\end{figure}

In our final set of additional merger simulations, we vary the gas fraction of
our fiducial G3 galaxy model.  To this end, the total mass in the disk is kept
fixed, yet the distribution of mass, i.e. the amount in the gaseous versus the
stellar disk, is varied for each model.  All other parameters, including the
interaction orbit and the satellite galaxies (G2, G1, and G0) remain unchanged
from their fiducial values.  Figure~\ref{fig:sfr_f} presents the star formation
history, the gas consumption, and the burst efficiency for the fiducial G3G2
interaction plus two additional mergers when the G3 primary has a 
larger gas fraction, $f$, defined as the mass of the gas disk divided by the 
total (gas plus stars) disk mass.

Figure~\ref{fig:sfr_f} exemplifies the generic outcome of all interactions
that test the gas fraction of the primary, namely, increasing the primary's
gas fraction decreases the burst efficiency.  In effect, the result is
similar to that presented in \S\ref{ssec:gasd} for the distribution of 
gas and follows from a similar cause --- the increased gas consumption
in the isolated primary.  In both scenarios, the initial disk converts a
large fraction its gas into stars regardless of whether the interaction
occurs or not.

\section{Discussion}
\label{sec:disc}

In this paper we perform a series of numerical simulations that follow the
interaction and merger of binary galaxies with various mass ratios.  Our analysis
quantifies the starbursts that result from the tidal forces that attend the
merger.  As expected, mergers between galaxies with nearly equal mass generate
the largest tidal forces, and therefore produce the most intense bursts of star
formation.  Mergers between galaxies with a large mass ratio produce relatively
little, if any, enhancement in star formation above quiescent evolution, yielding
a correlation between merger--driven star formation and decreasing merging galaxy
mass ratio.

To quantify the relationship between merging galaxy mass ratio and merger-driven
starburst, we introduce the burst efficiency (Eq.~\ref{eq:be}) as the fraction of gas
that is converted into stars during the interaction which does not occur if the
galaxies are evolved in isolation.  The burst efficiency is demonstrated
to be robust to uncertainties in the feedback parameterization unlike the strength
or duration of the starburst.

While the burst efficiency resulting from collisions between galaxies of equivalent
mass is relatively insensitive to the details of the merging event such as the
orbit, the galaxy orientation, and the properties of the merging galaxies \citep{
MH96,Sp00,Thesis}, this is not the case when the participating galaxies are unequal
in mass.

By performing a large number of additional merger simulations, we have 
quantified the effects of merging orbit and orientation, as well as properties of
the progenitor disk.  We find that close passage, co-planar orbits produce the
most significant bursts of star formation, consistent with the expectations of the
tidally induced origin for the starbursts.  We also find that the structure of the
progenitor disk strongly influences the merger-driven star formation.  In 
particular, the presence of a centrally-concentrated stellar bulge stabilizes the
disk and suppresses merger-driven star formation.  The distribution and mass of
the gaseous disk also influences the starburst.  In general, increasing the amount
of gas at densities above $\rho_{\rm crit}$, the threshold density for star formation
to commence, decreases the burst efficiency.

These results lead us to conclude the
following two facts about merger-driven starburst during an unequal mass galaxy merger:
(1) significant starbursts occur for only specialized scenarios, e.g., close passage,
co-planar orientations, when the primary disk does not contain a bulge,
and (2) even in this extreme scenario, the burst efficiency is still only $\lesssim0.25$,
i.e., a single unequal mass merger does not convert a large fraction of gas into stars.

\subsection{Comparison to Previous Simulations}
\label{ssec:compS}

Our work is closely related to, and consistent with, a number of prior studies of
the gas dynamics during the interaction and merger of unequal mass galaxies 
\citep[e.g.,][]{H89,MH94minm,HM95}.  While all work performed to date demonstrates
that minor mergers can induce radial inflows of gas that result in periods of
enhanced star formation, our models typically result in smaller burst efficiencies
(see \S\ref{sssec:sfloc} and \S\ref{sssec:be}) owing to three key differences 
in the studies.

First, as shown in \citet{Cox06} using simulations of major mergers, the more complex
treatment of the interstellar medium produces more hot gas, and suppresses star
formation.  In particular, relaxing the isothermal gas assumption commonly employed
in the work of Mihos \& Hernquist, including the ``conservative--entropy''
\citep{SHEnt} version of SPH, and the more efficient feedback models of \citet{Cox06}
all serve to suppress the merger--induced starburst, and lower the burst efficiency.

Second, the set of simulations employed by \citet{MH94minm} and \citet{HM95} followed
satellite galaxies that were initially placed on a circular orbit.  In contrast, we
follow parabolic orbits which are motivated by cosmological expectations (see 
\S\ref{sec:mergers}).  These more energetic orbits lead to less direct coupling between
the orbital angular momentum of the satellite and the disk of the primary, and
therefore are less conducive to an intense inflow of disk gas.

Lastly, as noted in \S\ref{ssec:bdrat} and in \citet{Cox06}, the star formation
model employed by Mihos \& Hernquist is less efficient than current observations
suggest \citep[e.g.,][]{Kenn98}.  The primary result of this assumption is that
the quiescent galaxy consumes much less gas when evolved in isolation and the merger
burst efficiency is overestimated.

While there are additional differences between the galaxy models employed by these
previous studies, namely they use less massive and less concentrated dark matter halos,
and the baryonic components are not as faithful a representation of observed galaxies
in the local Universe, the tests performed in \S\ref{sec:omods} indicate that these
differences play a secondary role to the three items outlined above.  Even when we
modified every possible parameter to maximize the burst efficiency, our value was
still only one--third as large as the previous results.

\subsection{Comparison to Observations}
\label{ssec:compO}

While observational studies have already established a clear link between star formation
and galaxy interactions \citep[as measured by close pairs or morphology, e.g.,][]
{LT78,JW85,Ken87,Bar03,Lam03,NCA04}, only recently has it been possible to
specifically address whether or not this correlation holds when the close galaxy
pairs have rather disparate luminosities.  In particular, \citet{WGB06} reported
no correlation between star formation (as measured by H$\alpha$) and close galaxy
pairs in the CfA2 Redshift Survey when the magnitude difference is greater than 2,
and a more recent study using a larger sample from SDSS found that the satellite
galaxy indeed shows enhanced star formation while the primary did not \citep{WG07}.

Assuming that luminosity traces mass (and modulo any systematic affects owing to 
merger--driven star formation), the studies by Woods et al. imply that there is no observational
evidence for induced star formation in the primary galaxy when the merger mass ratio is
greater than $\sim6$:1, while the satellite is more likely to experience a starburst.  Taken
at face value, our results naturally recover this observed trend.  Our merger models
produce little, if any, globally enhanced star formation for merger mass ratios below 5:1, and when they do,
it requires very specific circumstances (bulgeless primary, co--planar, close--passage
orbit).  It is also intriguing that Figure~\ref{fig:sfrg12} hints that the satellite
is much more susceptible to enhanced star--formation during the interaction, which also
seems consistent with the observations.

In contrast to the statistical studies, there are a multitude of
observations that suggest that individual systems are currently undergoing minor
merger induced episodes of star formation (see the list in \S\ref{sec:intro}).  To
determine whether these specific galaxies are consistent with the statistical
studies requires more extensive modeling of individual systems, as has been
performed in a few cases already \citep[see, e.g.,][]{MB97,LH99,SL00}.  It is
possible that many of these systems have unique satellite orbits, or that the effects
of multiple minor mergers which occur simultaneously is more dramatic than the binary
mergers that we have followed here.  Models of individual galaxies, and their direct
comparison to observations across many wavebands may also yield important constraints
to the star formation and feedback models.

\subsection{Implications for Dwarf Galaxies}
\label{ssec:dwarfs}

While the focus of our analysis has been on the global properties of 
merger-driven star formation, Figure~\ref{fig:sfrg12} clearly demonstrates
that the star formation history of the satellite galaxy can be enhanced
far more dramatically than that of the primary galaxy (which dominates the
global star formation).  Such a scenario appears to be observed in the large
sample of SDSS galaxies studied by \citet{WG07} and also in the nearby
Universe, e.g., both M82 and NGC 3077 are currently experiencing periods of
intense star formation after recent close passages to M81 \citep{Yun94,Wal02,
Ott03}.  These tidally--induced episodes of star formation may also have 
implications for the detection of satellite galaxies and the inferred
cosmological merger rate \citep[see, e.g.,][]{Ber06}.

\subsection{The Mass--Dependence of Star Formation}
\label{ssec:massdep}

One interesting feature present in Figure~\ref{fig:bestd} is the systematic
dependence of $e_{1:1}$, the burst efficiency for an equal--mass major merger, on
primary galaxy mass.  Specifically, the burst efficiency is 0.46 for the G3G3
major merger and steadily increases to 0.61 for the G0G0 major merger.

The increasing burst efficiency $e_{1:1}$ with decreasing system mass is a
direct byproduct of systematic changes in the merger--induced star formation
compared to that in the isolated disks.  One possibility for this trend is
the systematic increase in gas fraction with decreasing galaxy mass that is 
assumed for our galaxy models.  However, the results of  \S\ref{ssec:gasf}
suggest that increasing the gas fraction actually decreases the burst efficiency.

A more likely scenario is that the density--dependent description of star
formation, including the explicit density threshold $\rho_{\rm crit}$ for 
star formation to commence, is producing the large variation in burst 
efficiency with mass.  In particular, Figure~\ref{fig:sfriso} shows that the
quiescent star formation spans three orders of magnitude from G3 ($\sim1$
\msunyr) to G0 ($\sim10^{-3}$\msunyr), while the peak star formation during
the major mergers are much more comparable (G3G3 $\sim25$\msunyr and G0G0
$\sim3$\msunyr).  The significant increase in merger--induced star formation
compared to quiescent levels was noted in \S\ref{ssec:sfr} which determined
that the G3G3 merger enhances star formation by a mere order of magnitude
while the G0G0 merger enhances star formation by a whopping three orders of
magnitude.

More work is required to determine the precise nature of the mass--dependence
of star formation and how this trend depends upon uncertainties associated
with our implementation of star formation.  Moreover, a systematic comparison 
to the observed relationship between specific star formation rate and stellar
mass, i.e. downsizing \citep{Cow96,BE00,Kau03}, may elucidate the physical
mechanisms responsible.

\subsection{Input for Future Studies}
\label{ssec:feedsam}

\begin{table}
\begin{center}
\caption{
Compilation of best fit burst efficiency
(see \S\ref{sssec:be} and Eq.~\ref{eq:be})
parameters for various sets of simulations
employed in this paper.  All simulations
analyzed for these fits used the $n2med$
feedback model.
}
\begin{tabular}{lccl}
\hline
Section & $e_{1:1}$ & $\gamma$ & Comment \\
\hline
\hline
\ref{sssec:be}   & 0.55 &  0.69 &  fiducial series \\
\hline
\ref{ssec:morb} & 0.49 &  0.94 &  R$_{\rm peri}=1.7$~kpc \\
\ref{ssec:morb} & 0.49 &  0.82 &  R$_{\rm peri}=3.4$~kpc \\
\ref{ssec:morb} & 0.49 &  0.66 &  R$_{\rm peri}=6.8$~kpc \\
\ref{ssec:morb} & 0.50 &  0.74 &  R$_{\rm peri}=13.6$~kpc \\
\ref{ssec:morb} & 0.51 &  0.87 &  R$_{\rm peri}=27.2$~kpc \\
\ref{ssec:morb} & 0.50 &  0.96 &  R$_{\rm peri}=64.4$~kpc \\
\hline
\ref{ssec:morb} & 0.50 &  0.47 &  $0^\circ$ \\
\ref{ssec:morb} & 0.50 &  0.74 &  $30^\circ$ \\
\ref{ssec:morb} & 0.50 &  0.96 &  $60^\circ$ \\
\ref{ssec:morb} & 0.49 &  1.25 &  $90^\circ$ \\
\ref{ssec:morb} & 0.49 &  0.99 &  $150^\circ$ \\
\ref{ssec:morb} & 0.51 &  0.73 &  $180^\circ$ \\
\hline
\ref{ssec:bdrat} & 0.50 &  0.61 & B/D= 0.0 \\
\ref{ssec:bdrat} & 0.50 &  0.74 & B/D= 0.2 \\
\ref{ssec:bdrat} & 0.50 &  1.02 & B/D= 0.5 \\
\hline
\ref{ssec:gasf} & 0.50 &  0.74 & $f$= 0.20 \\
\ref{ssec:gasf} & 0.49 &  0.74 & $f$= 0.50 \\
\ref{ssec:gasf} & 0.44 &  0.72 & $f$= 0.78 \\
\hline
\end{tabular}
\label{tab:beps}
\end{center}
\end{table}

Quantifying the merger-driven star formation as a function of merger mass
ratio is useful for a variety of further studies.  In particular, semi-analytic
models of galaxy formation (SAMs) often find that merger-driven star formation
is necessary to reproduce the luminosity function and number counts of 
Lyman-break galaxies \citep{SPF} and sub-millimeter galaxies \citep{Gui98,
Bau05}.  To this end, we have followed the methodology of \citet{SPF} and 
introduced the burst efficiency (see \S\ref{sssec:be} and Eq.~\ref{eq:be}) in
order to parameterize the star formation enhancement that occurs during galaxy
interactions as a function of participant mass ratio.

In Table~\ref{tab:beps} we list the best fit parameters to Eq.\ref{eq:be}
for all models explored in this paper, including the fiducial series of
runs (\S\ref{sssec:be}), and all additional models (orbits in
\S\ref{ssec:morb}, bulge--to--disk ratios in \S\ref{ssec:bdrat}, and gas fractions
of the progenitor disk in \S\ref{ssec:gasf}) that use the G3 primary
galaxy.  A quick inspection of Table~\ref{tab:beps} indicates that the
parameter $e_{1:1}$, which normalizes Eq.~\ref{eq:be} to the burst efficiency
of a major merger, changes relatively little across this wide range of tests,
although it may depend on galaxy mass (see the discussion of \S\ref{ssec:massdep}).
On the other hand, the parameter $\gamma$, which determines the mass ratio
scaling, has a significant dependence on orbital inclination and bulge--to--disk
ratio, a mild dependence on the orbital angular momentum, and a negligible
dependence upon gas fraction.

While the burst efficiency is a useful quantification of the star--formation
induced by a galaxy merger, a better understanding of the star--formation
timescale is necessary to completely describe merger--induced star formation.
Unfortunately, as shown in \S\ref{sssec:sfts}, uncertainties in the feedback
model do not permit an unambiguous characterization of the star--formation
timescale.  There is hope, however, that better constraints can be placed
on the feedback model through more detailed modeling of individual systems
systems or by comparison to the observed distribution of star formation rates.

Finally, we emphasize that the accretion events followed in this study drive
structural evolution which is also an interesting and relevant input parameter
to future studies of galaxy formation and evolution.  In particular, the morphology
of the merger remnants (see Figs.~\ref{fig:G3G2stellarmorph}--\ref{fig:G3G1gasmorph})
suggests that the stellar and gaseous components react differently to the 
dynamical perturbation and that the remnant appears to be a systematically
earlier Hubble type than the original disk.  This is also consistent with
the surface density profile shown in Figure~\ref{fig:prof}, which shows an
excess of mass at small radii indicative of the formation of a stellar bulge.

\subsection{Other Considerations}
\label{ssec:other}

As a final comment, we note that the present simulations lack several
physical processes that may play a role in the results we have
presented.  First, we have not included the recycling of gas from
stellar winds and supernovae, an omission that would increase the gas
fraction as a function of time.  Another process that may increase the
gas content on long timescales is the accretion of gas from the
cosmological growth of structure.  While we eventually intend to
incorporate these effects into our modeling \citep[as has been done by
several authors already, see, e.g,][]{Tor04,CSca05,Stin06}, the
relatively small difference ($\sim15$\%) in the burst efficiency for
runs of different initial gas fractions (see \S\ref{ssec:gasf})
supports the notion that this omission has a relatively minor effect
on the star formation that results from a single merger.  However, for
the long--term evolution, and for the absolute star--formation rate,
these effects are likely to be important.

It should also be pointed out that our galaxy models are calibrated to
match low redshift observations of disk galaxies.  While we have surveyed
a small portion of the parameter space that might correspond to properties
of disks at higher redshift, e.g., higher gas fractions and more compact
initial disks, and therefore have some indication of the changes that
may occur, future studies will need to investigate these dependencies in
more depth.

We have also not included accreting black holes and their associated feedback
into the models presented here.  Recent work has shown that these processes may
play a significant role during major mergers, resulting in the formation of quasars
\citep{Hop06big}, and leading to remnants that reside on the M$_{\rm BH}-\sigma$
relation \citep{dMSH05,SdMH05,Rob06b} and have the colors \citep{SdMH05red} and other
properties \citep{Cox06x,Rob06fp} appropriate for present day elliptical galaxies.
While the minor interactions we discuss here are a likely fueling mechanism for
many forms of nuclear activity \citep[see, e.g.,][]{Hop06ll}, because significant
black hole growth requires large quantities of gas (approximately the entire 
content of the initial disks) to be driven to the galactic center, it is likely
that black hole growth and feedback play a relatively minor part during most of
the unequal mass interactions we follow here.  Of course, the existence of a
correlation between the black hole mass and the bulge mass \citep[see, e.g.,][]
{Mag98} suggests that at the very minimum the black hole imparts enough feedback
to regulate its own growth.

\section*{Acknowledgments}

We thank Gurtina Besla, Suvendra Dutta, Lars Hernquist, Phil Hopkins,
Jennifer Lotz, Greg Novak, Brant Robertson, and Josh Younger for
helpful discussions, and Eric Bell for providing his data in
electronic form.  This research used the Beowulf UpsAnd at UCSC and
computational resources at the National Energy Research Scientific
Computing Center (NERSC), which is supported by the Office of Science
of the US Department of Energy.  TJC, AD, and JRP acknowledge support
from NASA and NSF grants at UCSC.  PJ was supported by a UC/LLNL
cooperative grant from IGPP to Wil van Breugel, and by program numbers
HST-AR-10678 and HST-AR-10958, provided by NASA through grants from
the Space Telescope Science Institute, which is operated by the
Association of Universities for Research in Astronomy, Incorporated,
under NASA contract NAS5-26555.

\bibliographystyle{mn2e}
\bibliography{/home/tcox/Refs/refs}

\end{document}